\documentclass[aps,pre,10pt,showpacs,floatfix,onecolumn,nofootinbib,a4paper]{revtex4-2}

\usepackage{amssymb, amsmath, amsthm, mathtools}
\usepackage{caption}
\usepackage{natbib}
\usepackage{verbatim}
\usepackage{epsfig,color}
\usepackage{url}
\usepackage[english]{babel}

\usepackage{mathrsfs}
\usepackage{bm}
\usepackage{hyperref}
\usepackage{graphicx}
\usepackage{float}
\usepackage{subcaption}
\newtheorem{thm}{Theorem}
\newtheorem{definition}{Definition}

\usepackage{verbatim}

\newcommand{\tr}{\operatorname{tr}} 
\newcommand{\dd}{\operatorname{d}\!}
\newcommand{\system}[1]{\left\{\begin{aligned}#1\end{aligned}\right.}

\newcommand{\eref}{\eqref}
\newcommand{\eqalign}[1]{\begin{split}{#1}\end{split}}

\newcommand{\ack}[1]{\begin{acknowledgements}#1\end{acknowledgements}}

\newcommand{\R}{\mathbb{R}}
\newcommand{\Z}{\mathbb{Z}}

\newcommand{\sphere}{\mathbb{S}^2}
\newcommand{\surf}{\mathscr{S}}
\newcommand{\curve}{\mathscr{C}}

\newcommand{\great}{\curve_\beta}

\newcommand{\sech}{\operatorname{sech}}

\newcommand{\arctanh}{\operatorname{arctanh}}
\newcommand{\grad}{\nabla} 
\newcommand{\dv}{\operatorname{div}} 
\newcommand{\curl}{\operatorname{curl}} 
\newcommand{\gradn}{\grad\n}
\newcommand{\id}{{\mathbf{I}}}
\newcommand{\tP}{\mathbf{P}}
\newcommand{\prjn}{\tP(\n)}
\newcommand{\tW}{\mathbf{W}}
\newcommand{\skwn}{\mathbf{W}(\n)}
\newcommand{\splay}{S} 
\newcommand{\twist}{T} 
\newcommand{\bend}{\bm{b}} 
\newcommand{\osp}{q} 
\newcommand{\ospt}{\mathbf{D}} 
\newcommand{\dchar}{(\splay,\twist,b_1,b_2,\osp)}
\newcommand{\grads}{\nabla\!_\mathrm{s}} 
\newcommand{\dvs}{\operatorname{\dv_{s}}} 
\newcommand{\curls}{\operatorname{\curl_{s}}} 
\newcommand{\splays}{\splay_{\rm s}} 
\newcommand{\twists}{\twist_{\rm s}} 
\newcommand{\bends}{\bend_{\rm s}} 
\newcommand{\osps}{\osp_{\rm s}} 
\newcommand{\ospts}{\ospt_{\rm s}} 
\newcommand{\gradc}{\nabla\!_\mathrm{c}} 
\newcommand{\dvc}{\operatorname{\dv_{c}}} 
\newcommand{\curlc}{\operatorname{\curl_{c}}} 
\newcommand{\splayc}{\splay_{\rm c}} 
\newcommand{\twistc}{\twist_{\rm c}} 
\newcommand{\bendc}{\bend_{\rm c}} 
\newcommand{\ospc}{\osp_{\rm c}} 
\newcommand{\osptc}{\ospt_{\rm c}} 
\newcommand{\ellK}{\mathsf{K}}
\newcommand{\ellF}{\mathsf{F}}
\newcommand{\surface}{\mathscr{S}}

\newcommand{\frameC}{(\ex,\ey,\ez)}
\newcommand{\framee}{(\eu,\ev,\normal)}
\newcommand{\framen}{(\n,\np,\normal)}
\newcommand{\va}{\bm{a}}
\newcommand{\rv}{\bm{r}}
\newcommand{\wv}{\bm{w}}
\newcommand{\cv}{\bm{c}}
\newcommand{\dvv}{\bm{d}_v}
\newcommand{\dvu}{\bm{d}_u}
\newcommand{\normal}{\bm{\nu}}
\newcommand{\euclid}{\mathscr{E}}
\newcommand{\transl}{\mathscr{V}}
\newcommand{\W}{\mathbf{W}}
\newcommand{\curvature}{(\grads\normal)}
\newcommand{\skw}{\operatorname{skw}}
\newcommand{\sgn}{\operatorname{sgn}}
\newcommand{\nablastwo}{\nabla^2\!\!\!\!_\mathrm{s}\;}

\newcommand{\dphi}{\dot{\varphi}}
\newcommand{\dtheta}{\dot{\vartheta}}

\newcommand{\alphap}{\alpha_{,\varphi}}
\newcommand{\alphat}{\alpha_{,\vartheta}}
\newcommand{\au}{\alpha_{,u}}
\newcommand{\av}{\alpha_{,v}}

\newcommand{\n}{\bm{n}}
\newcommand{\np}{\bm{n}_{\perp}}
\newcommand{\dframe}{(\n_{1},\n_{2},\n)}
\newcommand{\ex}{\bm{e}_{x}}
\newcommand{\ey}{\bm{e}_{y}}
\newcommand{\ez}{\bm{e}_{z}}
\newcommand{\er}{\bm{e}_{r}}
\newcommand{\et}{\bm{e}_{\vartheta}}
\newcommand{\ep}{\bm{e}_{\varphi}}
\newcommand{\erd}{\dot{\bm{e}}_{r}}
\newcommand{\etd}{\dot{\bm{e}}_{\vartheta}}
\newcommand{\epd}{\dot{\bm{e}}_{\varphi}}

\newcommand{\rvu}{\rv_{,u}}
\newcommand{\rvv}{\rv_{,v}}
\newcommand{\eu}{\bm{e}_{u}}
\newcommand{\ev}{\bm{e}_{v}}
\newcommand{\vc}{\bm{c}}
\newcommand{\vd}{\bm{d}}
\newcommand{\vdu}{\bm{d}_u}
\newcommand{\vdv}{\bm{d}_v}
\newcommand{\vnu}{\bm{\nu}}
\newcommand{\vomega}{\bm{\omega}}
\newcommand{\vt}{\bm{t}}
\newcommand{\vtp}{\vt_\perp}
\newcommand{\vv}{\bm{v}}
\newcommand{\vp}{\bm{p}}
\newcommand{\vpd}{\dot{\bm{p}}}
\newcommand{\bp}{b_\perp}
\newcommand{\bn}{b_\nu}

\begin{document}
\latintext
\title{Surface  Nematic Quasi-Uniformity}
\author{Andrea Pedrini}
\email{andrea.pedrini@unipv.it}
\affiliation{Dipartimento di Matematica, Universit\`a di Pavia, Via Ferrata 5, 27100 Pavia, Italy}
\author{Epifanio G. Virga}
\email{eg.virga@unipv.it}
\affiliation{Dipartimento di Matematica, Universit\`a di Pavia, Via Ferrata 5, 27100 Pavia, Italy}
\date{\today}


\begin{abstract}
Line fields on surfaces are a means to describe the \emph{nematic} order that may pattern them. The least distorted nematic fields are called \emph{uniform}, but they can only exist on surfaces with negative constant  Gaussian curvature. To identify the least distorted nematic fields on a generic surface, we relax the notion  of uniformity into that of \emph{quasi-uniformity} and prove that all such fields are parallel transported (in Levi-Civita's sense) by the geodesics of the surface. Both global and local constructions of quasi-uniform fields are presented to illustrate both richness and significance of the proposed notion.
\end{abstract}

\maketitle

\section{Introduction}
Surfaces with a nematic order imprinted on them promise to be relevant to disparate soft matter systems. These include, for example, thin sheets of liquid crystal elastomers (see \cite{lagerwall:liquid} for an inspiring recent review and \cite{korley:introduction} for the introduction to a special journal's issue collecting many applications) and nematic shells, colloidal particles (or droplets) coated by a thin liquid crystal film (the relevant literature is vast, we only cite the recent contributions \cite{napoli:nematic,he:from} as possible sources for  the interested reader).

At the foundation of a general elastic theory for the nematic order described by a director field $\n$ must lie the notion of \emph{ground state}, as the less distorted arrangement of directors. If \emph{uniformity} has proven to be the right theoretical tool to handle this issue in three-dimensional space, it generally fails on surfaces, as it was proved by Niv \& Efrati~\cite{niv:geometric} that uniform fields can only exist on surfaces with \emph{negative} constant Gaussian curvature.

Thus, we need a relaxed notion of uniformity, which we found in the concept of \emph{quasi-uniformity} we started exploring in \cite{pedrini:relieving}. Here, we conduct a systematic study and prove a theorem that characterizes all quasi-uniform nematic distortions on a surface. This theorem extends the one proved in \cite{pedrini:surface} for surface uniform fields, showing how both uniformity and quasi-uniformity of surface nematic fields share one and the same geometric structure.

The paper is organized as follows. In Section~\ref{sec:distortion}, we recall the invariant measures of distortion for a nematic field in three space dimensions, and we see in Section~\ref{sec:distortion_lower} how these are adapted to surfaces and curves. In Section~\ref{sec:geodesic}, we prove our main theorem, and we derive in Section~\ref{sec:character} the equation that governs the evolution on surfaces of a scalar function attached to a quasi-uniform distortion, which we call its \emph{character}. Section~\ref{sec:applications} illustrates a number of applications of the geometric method to construct quasi-uniform nematic fields on surfaces that emerges from our theorem: we contemplate both \emph{global} fields defined on the whole surface and \emph{local} extensions of a field prescribed uniformly on a curve. Finally, in Section~\ref{sec:colclusion}, we collect the conclusions of this study, and in the closing three appendices we gather a number of auxiliary results, which we deem useful for the reader caring about details.

\section{Distortion characteristics  in three space dimensions}\label{sec:distortion}
In three-dimensional Euclidean space  $\euclid$, endowed with translation space $\transl$ of vectors,\footnote{Our notation for $\euclid$ and $\transl$ is the same as in \cite[p.\,324]{truesdell:first}, where these geometric structures are further illuminated, especially in connection with the  theoretical foundations of modern continuum mechanics.} the spatial gradient $\gradn$ of a director field $\n$ into the unit sphere $\sphere$  encodes the distortion of $\n$ in the invariant measures of \emph{splay} $\splay:=\dv\n$, \emph{twist} $\twist:=\n\cdot\curl\n$, and \emph{bend} $\bend:=\n\times\curl\n$ extracted from  the following decomposition, first proposed in \cite{machon:umbilic} and then reinterpreted in \cite{selinger:interpretation},
\begin{equation}\label{eq:gradn_decomposition}
\gradn = -\bend\otimes\n + \frac12\twist\skwn + \frac12\splay\prjn + \ospt.
\end{equation}
In \eref{eq:gradn_decomposition}, $\skwn$ is  the skew-symmetric tensor  with axial vector $\n$,\footnote{A skew-symmetric tensor $\W(\bm{a})$ with axial vector $\bm{a}$ acts on any vector $\vv\in\transl$ as follows, $\W(\bm{a})\vv=\bm{a}\times\vv$, where $\times$ denotes the cross product of vectors.} $\prjn:=\id-\n\otimes\n$ is the projection  onto the plane  orthogonal to $\n$, $\id$ is the identity tensor, and $\ospt$ is a traceless symmetric tensor satisfying $\ospt\n=\bm{0}$. Whenever $\ospt\neq\bm{0}$, it can be represented in a diagonal form by conventionally selecting its positive eigenvalue $\osp>0$ and an orthonormal basis of eigenvectors $(\n_1,\n_2)$ orthogonal to $\n$ and such that $\n=\n_1\times\n_2$,
\begin{equation}\label{eq:ospt}
\ospt = \osp(\n_1\otimes\n_1 - \n_2\otimes\n_2).
\end{equation}
We call $\osp$ the \emph{octupolar splay}:\footnote{We refer the reader to \cite{pedrini:liquid} for a \emph{visual} justification of this name (see also \cite{gaeta:review}); other authors prefer to call $\osp$ the \emph{tetrahedral} splay \cite{selinger:director}.} it is the fourth invariant measure of distortion and, by \ref{eq:gradn_decomposition}, it can be given the form
\begin{equation}\label{eq:osp}
q = \frac1{\sqrt2}\sqrt{\tr(\gradn)^2 + \frac12\twist^2 - \frac12\splay^2}.
\end{equation}
In the \emph{distortion frame} $\dframe$, $\skwn$ and $\bend$ read as
\begin{equation}
\skwn = \n_2\otimes\n_1 - \n_1\otimes\n_2
\quad\mathrm{and}\quad
\bend = b_1\n_1 + b_2\n_2,
\end{equation}
for suitable scalars $b_1$ and $b_2$.\footnote{Whenever $\ospt=\bm{0}$, we conventionally define the distortion frame by choosing $\n_1:=\frac{\bend}{\|\bend\|}$ and $\n_2:=\n\times\n_1$, so that $\bend=b_1\n_1$ with $b_1=\|\bend\|$. If also $\bend=\bm{0}$, any orthonormal basis $(\n_1,\n_2)$ orthogonal to $\n$ can be chosen, with no loss of generality.} Hence, the distortion of $\n$ is described by the \emph{five} scalars $\dchar$; they are referred to as the \emph{distortion characteristics} of $\n$ (in three space dimensions).

Whenever the distortion characteristics are constant in the whole domain where $\n$ is defined, the representation of $\gradn$ in the intrinsic frame is everywhere the same and the distortion is called \emph{uniform}. As proved in \cite{virga:uniform}, the unique uniform distortions in the whole three-dimensional Euclidean space are Meyer's \emph{heliconical} distortions \cite{meyer:structural}, for which either
\begin{equation}\label{eq:heliconical}
S=0,\ T=2q,\ b_1=b_2
\quad\mathrm{or}\quad
S=0,\ T=-2q,\ b_1=-b_2.
\end{equation}

A natural way to relax  uniformity is to consider \emph{quasi-uniform} distortions, for which the distortion characteristics are in constant ratios to one another, rather than being constant themselves. The notion of quasi-uniformity, proposed in \cite{pedrini:liquid} and further addressed in \cite{pollard:intrinsic} for three-dimensional Euclidean space, has also been studied in \cite{pedrini:relieving} as an effective means to relieve one-dimensional nematic geometric frustration into  a half-plane. Both uniformity and quasi-uniformity are notions that can  be extended to lower dimensions, contingent upon the appropriate choice of distortion characteristics. In the following section, we shall digress slightly to make this choice precise.

\section{Distortion characteristics in lower dimensions}\label{sec:distortion_lower}
As the characteristics of distortion are invariant measures with an intrinsic meaning, independent of the observer and describing solely the  properties of the director field, the most suitable operators to extract them in lower dimensions are the covariant gradient and derivative of $\n$. This approach naturally disregards any component that depends extrinsically only on the domain's curvature.
\begin{figure}[h!]
\centering
\begin{subfigure}[b]{0.45\textwidth}
\centering
\includegraphics[width=0.9\textwidth]{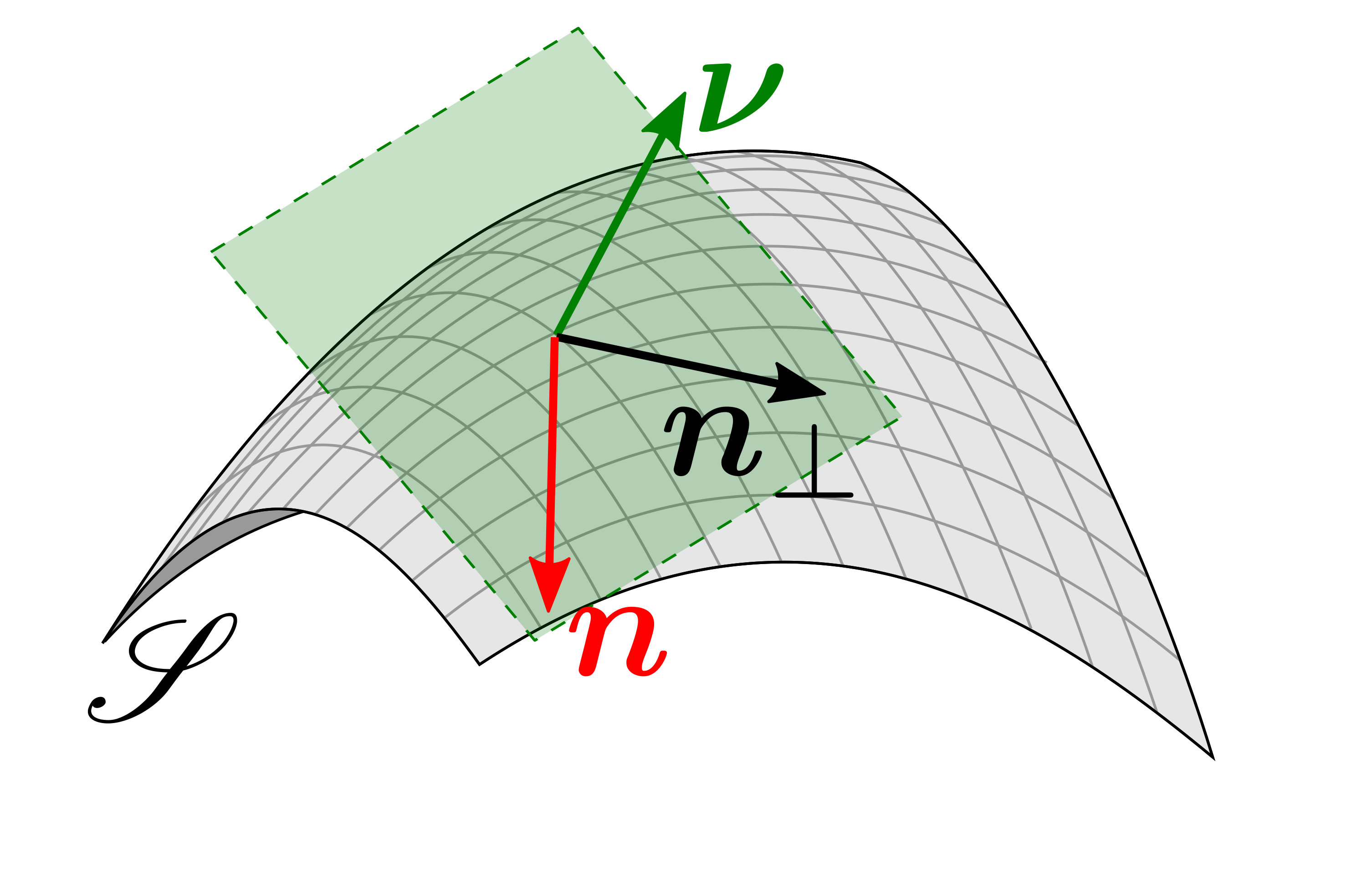}
\caption{Distortion frame gliding on the surface $\surf$.}\label{fig:local}
\end{subfigure}
$\quad$
\begin{subfigure}[b]{0.45\textwidth}
\centering
\includegraphics[width=0.9\textwidth]{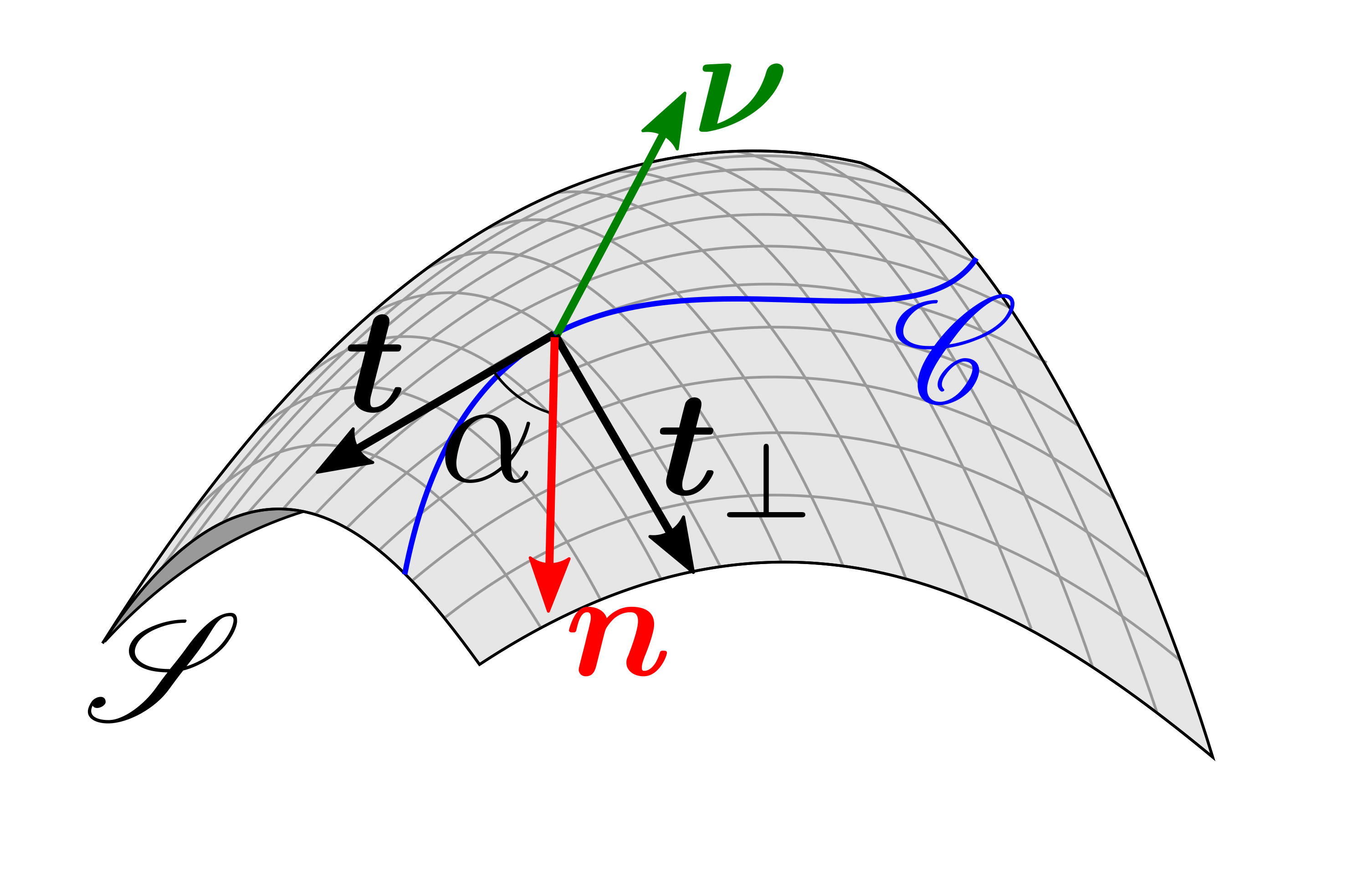}
\caption{Darboux frame moving along a curve $\curve$ on $\surface$.}\label{fig:darboux}
\end{subfigure}
\caption{Frames employed to represent the covariant gradient and derivative of $\n$.}
\end{figure}

\subsection{Surface quasi-uniformity}
Let $\surface$ be a regular surface (at least of class $C^2$) in three-dimensional space $\euclid$ with unit normal $\normal$ and let
$\n:\surf\to\sphere$ be a (locally twice) differentiable unit vector field everywhere tangent to $\surface$. Setting $\np:=\normal\times\n$, we define on $\surface$  the orthonormal frame $\framen$, which is the \emph{distortion} frame adapted to a surface (see Figure~\ref{fig:local}). Any smooth extension $\hat{\n}$ of $\n$ defined in a three-dimensional neighborhood of $\surf$ would serve the purpose of defining the \emph{surface} gradient $\grads\n$ as
\begin{equation}
\label{eq:surf_grad}
\grads\n:=(\grad\hat{\n})\tP(\vnu),
\end{equation}
where $\tP(\vnu):=\id-\vnu\otimes\vnu$ is the projection onto the local tangent plane of $\surface$. The \emph{surface} divergence and the \emph{surface} curl of $\n$ are then defined as\footnote{In particular, \eref{eq:curls} says that $\curls\n=2\bm{w}$, where $\bm{w}$ is the axial vector of the skew-symmetric part of $\grads\n$. Here and in what follows a superscript $^\top$ denotes transposition.}
\begin{eqnarray}
&\dvs\n := \tr\grads\n, \label{eq:divs} \\
&\curls\n\times\vv := \big[\grads\n - (\grads\n)^\top\big]\vv \quad \forall\ \vv\in\transl \label{eq:curls}.
\end{eqnarray}

Letting 
\begin{equation}
\label{eq:b_S_T_surface}	
\bends:=\n\times\curls\n,\quad\splays:=\dvs\n,\quad\textrm{and}\quad\twists:=\n\cdot\curls\n
\end{equation}
be the \emph{surface} bend, splay, and twist, respectively, as shown in Appendix~\ref{app:surface_gradient}, we can represent  $\grads\n$ in the following form
\begin{equation}
	\label{eq:surface_gradient_representation}
	\grads\n=-\bends\otimes\n+\splays\np\otimes\np+\twists\normal\otimes\np.
\end{equation}
It is an immediate consequence of \eref{eq:surface_gradient_representation} that $\grads\n$ also admits a decomposition similar to \eref{eq:gradn_decomposition}:
\begin{equation}\label{eq:surf_dec}
\grads\n = - \bends\otimes\n + \frac\splays2\prjn + \frac\twists2\skwn + \ospts,
\end{equation}
where  the tensors $\prjn$, $\skwn$, and $\ospts$ are given by
\begin{equation} \label{eq:local_skw}
\prjn=\np\otimes\np+\normal\otimes\normal,\quad \skwn = \vnu\otimes\np - \np\otimes\vnu,
\end{equation}
and 
\begin{equation}\label{eq:D_s}
\ospts := \frac12\big[\splays(\np\otimes\np - \vnu\otimes\vnu) + \twists(\vnu\otimes\np + \np\otimes\vnu)\big],
\end{equation}
respectively.
The latter is a symmetric and traceless tensor which annihilates $\n$, and whose non-negative eigenvalue is the \emph{surface} octupolar splay
\begin{equation}
\osps := \frac12\sqrt{\splays^2 + \twists^2}.
\end{equation}
To cast \eref{eq:D_s} in a form akin to \eref{eq:ospt}, it suffices to set
\begin{equation}
	\label{eq:D_s_akin}
	\n_1=\cos\theta\np+\sin\theta\normal,\quad\n_2=\cos\theta\normal-\sin\theta\np\quad \textrm{with}\ \tan2\theta=\frac{\twists}{\splays}.
\end{equation}
It easily follows from the definition of $\bends$ in \eref{eq:b_S_T_surface} and the symmetries enjoyed by the mixed product $\bm{a}\cdot\bm{b}\times\cv$ of vectors $\bm{a}$, $\bm{b}$, $\cv$ (see \eref{eq:mixed_product_symmetries} in Appendix~\ref{app:surface_gradient}) that
\begin{equation}\label{eq:bends}
\bends = \bp\np + \bn\vnu, 
\quad\textrm{where}\quad
\bp := -\vnu\cdot\curls\n\quad
\textrm{and}\quad
\bn := \np\cdot\curls\n.
\end{equation}

Projecting $\grads\n$ onto the plane orthogonal to $\vnu$ produces the \emph{covariant} gradient $\gradc\n$, which is \emph{intrinsic} to the surface $\surface$, insensitive to the imbedding of $\surface$ in the surrounding three-dimensional space $\euclid$ (see also \cite[p.\,240]{needham:visual}),\footnote{By contrast, properties of a surface $\surface$ that depend on the imbedding of $\surface$ in $\euclid$ are often called \emph{extrinsic}. Any property featuring the normal $\normal$ is likely to be extrinsic.}
\begin{equation}\label{eq:gradc}
\gradc\n := \tP(\vnu)\grads\n = -\bp\np\otimes\n + \splays\np\otimes\np,
\end{equation}
which follows at once from both \eref{eq:surface_gradient_representation} and \eref{eq:bends}.
By defining the \emph{covariant} bend, splay, and twist in complete analogy with \eref{eq:b_S_T_surface},
\begin{equation}
	\label{eq:b_S_T_covariant}
	\bendc:=\n\times\curlc\n,\quad\splayc:=\dvc\n,\quad\textrm{and}\quad\twistc:=\n\cdot\curlc\n,
\end{equation}
where, mimicking \eref{eq:divs} and \eref{eq:curls}, we have set
\begin{eqnarray}
	&\dvc\n := \tr\gradc\n, \label{eq:divc} \\
	&\curlc\n\times\vv := \big[\gradc\n - (\gradc\n)^\top\big]\vv \quad \forall\ \vv\in\transl \label{eq:curlc},
\end{eqnarray}
we easily obtain from \eref{eq:gradc} that
\begin{equation}\label{eq:b_S_T_covariant_computed}
	\bendc =\bp\np,\quad\splayc =\splays,\quad\textrm{and}\quad\twistc = 0.
\end{equation}

As shown in Appendix~\ref{app:surface_gradient}, $\curls\n$ can be decomposed as follows in the frame $\framen$,
\begin{equation}
	\label{eq:surface_curl_decomposition}
	\curls\n=-[\np\cdot\curvature\n]\n+[\n\cdot\curvature\n]\np+[\np\cdot(\grads\n)\n]\normal,
\end{equation}
where $\grads\normal$ is the \emph{curvature} tensor. Thus, by \eref{eq:bends}, we see that 
\begin{equation}
	\label{eq:b_T_surface_covariant}
	\bends=\bendc+[\n\cdot\curvature\n]\normal\quad\textrm{and}\quad\twists=-\np\cdot\curvature\n,
\end{equation}
which show that $\bends$ differs from $\bendc$ by an extrinsic vector, whereas $\twists$ is totally extrinsic. Accordingly, $b_\perp=-\np\cdot(\grads\n)\n$ and $\splayc=\splays$ emerge as the only \emph{intrinsic} surface distortion characteristics. Moreover, by use of \eref{eq:surface_curl_decomposition}, \eref{eq:bends}, and \eref{eq:surface_gradient_representation}, we easily arrive at
\begin{equation}
	\label{eq:surface_covariant_gradient}
	\grads\n=\gradc\n-\normal\otimes\curvature\n,
\end{equation}
which will play a role in Section~\ref{sec:character} below. Finally, \eref{eq:gradc} can also be recast in a form akin to \eref{eq:surf_dec},
\begin{equation}\label{eq:cov_dec}
	\gradc\n 
	= -\bendc\otimes\n + \frac\splayc2\tP(\n) + \osptc,
\end{equation}
where
\begin{equation}\label{eq:D_c}
	\osptc := \frac{\splayc}{2}(\np\otimes\np - \vnu\otimes\vnu),
\end{equation}
which leads us to identify the \emph{covariant} octupolar splay as $\ospc:=\frac{|\splayc|}2$.

Hereafter,  with no loss of generality, we shall simply denote by $S$ the common value of $\splays$ and $\splayc$,\footnote{No confusion should arise with the three-dimensional splay in \eref{eq:gradn_decomposition}, as this paper is entirely concerned with nematic fields on surfaces.} and so we rewrite \eref{eq:gradc} as
\begin{equation}
	\label{eq:covariant_gradient_representation}
	\gradc\n=-b_\perp\np\otimes\n+S\np\otimes\np,
\end{equation}
which, combined with \eref{eq:surface_covariant_gradient}, also delivers
\begin{equation}
	\label{eq:surface_gradient_combined}
	\grads\n=-b_\perp\np\otimes\n+S\np\otimes\np-\normal\otimes\curvature\n.
\end{equation}
\begin{definition}
	\label{def:quasi_uniformity}
	We say that a smooth director field $\n$ everywhere tangent to $\surface$ is a \emph{quasi-uniform} distortion if its intrinsic distortion characteristics are in constant ratio.
\end{definition}
Specifically, for a quasi-uniform distortion we can write
\begin{equation}\label{eq:quasi_uniformity_definition}
	b_\perp=BS,
\end{equation}
for some \emph{constant} $B\in\R$. We shall also refer to $B$ as the distortion \emph{anisotropy}. Letting $B_0$ be a reference (constant) bend and $S_0$ a reference (constant) splay, for a quasi-uniform distortion we can also set
\begin{equation}
	\label{eq:character_definition}
	b_\perp=fB_0\quad\textrm{and}\quad\splay=f\splay_0,
\end{equation}
where $f$ is a scalar-valued function defined on $\surface$, which we call the quasi-uniformity \emph{character} of $\n$. Requisites \eref{eq:quasi_uniformity_definition} and \eref{eq:character_definition} agree if the distortion anisotropy is expressed as
\begin{equation}
	\label{eq:B_formula}
	B=\frac{B_0}{S_0}.
\end{equation}
With no loss of generality, we can rescale lengths in $\euclid$ so that 
\begin{equation}
	\label{eq:scaling_identity}
	S_0^2+B_0^2=1,
\end{equation}
without affecting \eref{eq:B_formula}.

We finally remark that the \emph{uniform} distortions on $\surf$  studied in \cite{pedrini:surface} are quasi-uniform distortions for which both intrinsic distortion characteristics are constant everywhere on $\surf$. They have character $f\equiv\pm1$.

\subsection{Uniform transport}
As suggested in \cite{pedrini:relieving}, a director field $\n$ prescribed on a smooth curve $\curve$ represents a geometric frustration that can be relieved quasi-uniformly through a purely geometric mechanism, independent of energy considerations. Accordingly, we define here the concept of uniformity on curves, in a manner consistent with the notion of two-dimensional quasi-uniformity.

The \emph{Darboux frame} is a local frame $(\vt,\vtp,\vnu)$ along a curve $\curve$ of $\surf$, where $\vt$ is the unit vector tangent to $\curve$, $\vnu$ is the unit vector orthogonal to $\surf$ and $\vtp:=\vnu\times\vt$ (as illustrated in Figure \ref{fig:darboux}). A director field $\n$ defined on $\surf$ can be parametrized along $\curve$ using the arc-length parameter $\sigma$, oriented coherently with $\vt$,
\begin{equation}\label{eq:n_1d}
\n = \cos\gamma\vt + \sin\gamma\vtp \quad\textrm{with}\quad\gamma = \gamma(\sigma).
\end{equation}
As a consequence of \eref{eq:n_1d}, $\np = \vnu\times\n$ reads as
\begin{equation}\label{eq:n_perp_1d}
\np = - \sin\gamma\vt + \cos\gamma\vtp.
\end{equation}
Letting $'$ denote differentiation with respect to $\sigma$, the evolution of the Darboux frame along the curve $\curve$ is governed by the \emph{spin} vector $\vomega := \tau_g\vt - \kappa_n\vtp + \kappa_g\vnu$ (see, for example, \cite[p.\,264]{doCarmo:differential}), so that
\begin{equation}\label{eq:Darboux_gliding_equations}
\system{
&\vt' = \vomega\times\vt = \kappa_g\vtp + \kappa_n\vnu, \\
&\vtp' = \vomega\times\vtp = - \kappa_g\vt + \tau_g\vnu,\\
&\vnu' = \vomega\times\vnu = - \kappa_n\vt - \tau_g\vtp,
}
\end{equation}
where $\kappa_g$, $\kappa_n$, and $\tau_g$ are the \emph{geodesic curvature}, the \emph{normal curvature}, and the \emph{geodesic torsion} of $\curve$, respectively. By combining \eref{eq:n_1d}, \eref{eq:n_perp_1d}, and \eref{eq:Darboux_gliding_equations}, we can easily differentiate $\n$ along $\curve$,
\begin{equation}\label{eq:n_prime}
\n' = (\gamma' + \kappa_g)\np + (\tau_g\sin\gamma + \kappa_n\cos\gamma)\vnu.
\end{equation}

Since $\vnu'=(\grads\vnu)\vt$, it follows from the identity
\begin{equation}\label{eq:swap}
(\grads\n)^\top\vnu = -(\grads\vnu)\n,
\end{equation}
proved in Appendix~\ref{app:surface_gradient}, that 
\begin{equation}\label{eq:n_prime_bis}
\n'\cdot\vnu
= (\grads\n)\vt\cdot\vnu 
= - \vt\cdot(\grads\vnu)\n
= -\vnu'\cdot\n,
\end{equation} 
which neatly identifies $\n'\cdot\np=\gamma' + \kappa_g$ as the only intrinsic component of $\n'$. Using \eref{eq:covariant_gradient_representation}, we can also define the \emph{covariant} derivative $\n'_\mathrm{c}$ of $\n$ along $\curve$ as $\n'_\mathrm{c} :=(\gradc\n)\vt$, thus arriving at
\begin{equation}\label{eq:nprime_cov}
\n'_\mathrm{c} := - (\bp\cos\gamma + \splay\sin\gamma)\np,
\end{equation}
also by use of \eref{eq:n_1d} and \eref{eq:n_perp_1d}. By requiring that $\n'_\mathrm{c}\cdot\np=\n'\cdot\np$, we then obtain an equation for $\gamma$ on $\curve$,
\begin{equation}\label{eq:gamma_equation}
\gamma'+\kappa_g+\bp\cos\gamma + \splay\sin\gamma=0,
\end{equation}
which expresses compatibility between a director field on $\surface$ and its \emph{trace} along $\curve$.

\begin{definition}
	\label{def:curve_uniformity}
	We say that a director field $\n$ is \emph{uniform} along a curve $\curve$ on $\surface$ if
	\begin{equation}
		\label{eq:curve_uniformity}
	\n'\cdot\np=\n'_\mathrm{c}\cdot\np\equiv g,
	\end{equation}
where $g\in\R$ is constant.\footnote{Since along a curve $\curve$ there is a single intrinsic measure of distortion, there is no room to define quasi-uniformity on it.}
\end{definition}
It follows from \eref{eq:n_prime_bis} that \eref{eq:curve_uniformity} amounts to require that $\gamma$ solves the equation
\begin{equation}
	\label{eq:gamma_equation_bis}
	\gamma'+\kappa_g=g.
\end{equation}
Letting $\gamma_0:=\gamma(0)$ be prescribed, we obtain from \eref{eq:gamma_equation_bis} that
\begin{equation}\label{eq:alpha_kappa}
\gamma(\sigma) = \gamma_0 + g\sigma - \int_0^\sigma\kappa_g(\tau)\dd\tau,
\end{equation}
indicating that on any regular curve $\curve$ there exist infinitely many uniform distortions, each associated with a distinct choice of the constants $\gamma_0$ and $g$. The same rich variety of possibilities arises when the curve is closed; however, in this case, the following compatibility condition must be imposed for $\n$ to be continuous on $\curve$, to within the nematic symmetry that identifies $\n$ and $-\n$:
\begin{equation}\label{eq:closure_condition}
\gamma(\ell) -\gamma_0=k\pi,
\quad\textrm{with }k\in\Z,
\end{equation}
where  $\ell$ is the length of $\curve$. By combining \eref{eq:alpha_kappa} and \eref{eq:closure_condition}, $g$ is determined as
\begin{equation}
g = \frac1\ell\Bigg[k\pi + \int_0^\ell\kappa_g(\tau)\dd\tau\Bigg],
\end{equation}
while both $\gamma_0$ and $k$ remain free.

It is worth noting that for $g=0$ equation \eref{eq:gamma_equation_bis} reduces to the condition that defines Levi-Civita's \emph{parallel transport} along a curve on a surface, for which $\n'_\mathrm{c}\equiv\bm{0}$ (see \cite{levi-civita:nozione} for a general definition and \cite{persico:realizzazione} for a kinematic interpretation applicable to our setting). Thus, we may also say that \eref{eq:gamma_equation_bis}, along with \eref{eq:n_1d} and \eref{eq:alpha_kappa}, define a \emph{uniform transport} of $\n$ along $\curve$.

A question then arises naturally: whether distortions realized by a uniform transport along a curve $\curve$ can be extended quasi-uniformly (and possibly globally) across the entire hosting surface $\surf$. This issue will be addressed in Section~\ref{sec:applications}.

In the next section, we delve deeper into the relation between surface quasi-uniformity and parallel transport, proving that quasi-uniform distortions arise uniquely from the parallel transport of $\n$ along geodesics. 

\section{Characteristic geodesics}\label{sec:geodesic}
Let $\rv:(u,v)\mapsto\rv(u,v)$ be an \emph{isothermal} parametrization  of a
regular surface $\surf$,\footnote{The existence of isothermal coordinates for surfaces of class $C^2$ is a classical results revisited in \cite{chern:elementary}.} so that
\begin{equation}
 \|\rvu\| = \|\rvv\|=:r\quad
\textrm{and}\quad
\rvu\cdot\rvv = 0,\quad
\textrm{where}\quad
\rvu := \frac{\partial\rv}{\partial u},\ 
\rvv := \frac{\partial\rv}{\partial v},
\end{equation}
and let $\framee$ be the moving frame associated with the isothermal coordinates $(u,v)$,\footnote{The method of \emph{moving} frames was introduced by Cartan \cite{cartan:methode}; a more recent account can be found in \cite{clelland:from}. The reader is also referred to \cite{o'neill:elementary} for the application of this method to the differential geometry of curves and surfaces.}
\begin{equation}\label{eq:isothermal_frame}
\eu:=\frac{\rvu}r,\quad
\ev:=\frac{\rvv}r,\quad
\vnu=\eu\times\ev.
\end{equation}
The motion of the frame $\framee$ is governed by the following \emph{gliding laws} (see also \cite{pedrini:surface}), 
\begin{equation}\label{eq:gliding_laws}
\system{
&\grads\eu = \ev\otimes\vc + \vnu\otimes\vdu, \\
&\grads\ev = -\eu\otimes\vc + \vnu\otimes\vdv, \\
&\grads\vnu = -\eu\otimes\vdu - \ev\otimes\vdv,
}
\end{equation}
where the vector fields $(\vc,\vdu,\vdv)$ are everywhere tangent to $\surface$; these are the \emph{connectors} of the frame $\framee$: $\vc$ is the \emph{spin} connector and $\vdu$, $\vdv$ are the \emph{curvature} connectors. Since the curvature tensor $\grads\normal$ is symmetric, the curvature connectors must obey the identity
\begin{equation}
	\label{eq:connector_identity}
	\vdv\cdot\eu=\vdv\cdot\eu.
\end{equation}
Moreover, the third equation in \eref{eq:gliding_laws} implies that 
\begin{eqnarray}
	&2H:=\tr\curvature=-(\vdu\cdot\eu+\vdv\cdot\ev),\label{eq:H}\\
	&K:=\det\curvature=\vdu\times\vdv\cdot\normal,\label{eq:K}
\end{eqnarray}
where $H$ and $K$ are the \emph{mean} and \emph{Gaussian} curvatures of $\surface$, respectively.\footnote{Here, $\tr\curvature$ and $\det\curvature$ are the sum and product of the principal curvatures of $\surface$, respectively.}

In the frame $\framee$, we represent the director field $\n$ as
\begin{equation}\label{eq:n_representation}
\n = \cos\alpha\eu + \sin\alpha\ev,
\end{equation}
where $\alpha = \alpha(u,v)$ is a differentiable function of the coordinates.
By letting $\vc=c_u\eu+c_v\ev$, since by \eref{eq:n_representation} $\np =\normal\times\n= -\sin\alpha\eu + \cos\alpha\ev$, a straightforward chain of calculations employing \eref{eq:gliding_laws} and \eref{eq:isothermal_frame} leads us to
\begin{equation}
\eqalign{
\grads\n &=
\np\otimes\grads\alpha
+\cos\alpha\grads\eu + \sin\alpha\grads\ev \\
&=\frac\np r\otimes(\au\eu+\av\ev) 
+ \np\otimes\vc 
+ \vnu\otimes(\cos\alpha\vdu+\sin\alpha\vdv),
}
\end{equation}
whence, also with the aid of \eref{eq:np_gradn_np} and \eref{eq:curlsn_nu}, we arrive at
\begin{eqnarray}
&\splay = \dvs\n = \Big(\frac{\av}r+c_v\Big)\cos\alpha - \Big(\frac{\au}r+c_u\Big)\sin\alpha, \\
&\bp = -\vnu\cdot\curls\n
= - \Big(\frac{\av}r + c_v\Big)\sin\alpha - \Big(\frac{\au}r + c_u\Big)\cos\alpha.
\end{eqnarray}
Thus, the requirement of quasi-uniformity in \eref{eq:quasi_uniformity_definition} becomes the following quasi-linear partial differential equation for $\alpha$,\footnote{The reader is reminded that partial differentiation is denoted by appending a comma.} 
\begin{equation}\label{eq:quasi_pde}
(\cos\alpha-B\sin\alpha)\frac{\au}r + (\sin\alpha+B\cos\alpha)\frac{\av}r 
= -(\cos\alpha-B\sin\alpha)c_u - (\sin\alpha+B\cos\alpha)c_v,
\end{equation}
where  $B$ is a constant.\footnote{It is perhaps worth remarking that the components  $c_u$ and $c_v$ of the spin connector $\vc$ are functions of $(u,v)$ depending on the choice of isothermal coordinates.}

\subsection{Characteristic lines}
To solve \eref{eq:quasi_pde}, we apply the method of \emph{characteristics}. These are lines in the $(u,v,\alpha)$ space that generate the integral surfaces of \eref{eq:quasi_pde} (see, e.g., \cite[p.\,62-69]{courant:methods} or \cite[Sec.~4.8]{salsa:partial}). They obey the following system of ordinary differential equations in the auxiliary parameter $t\in\R$,\footnote{This is also called the Lagrange-Charpit system \cite{delgado:lagrange}.}
\begin{equation}\label{eq:lagrange_charpit}
\system{
&\dot{u} = \frac1r(\cos\alpha-B\sin\alpha),\\
&\dot{v}  = \frac1r(\sin\alpha+B\cos\alpha), \\
&\dot\alpha= -(\cos\alpha-B\sin\alpha)c_u - (\sin\alpha+B\cos\alpha)c_v,
}
\end{equation}  
where a superimposed dot $\dot{\null}$ denotes differentiation with respect to $t$.
The projection on the $(u,v)$ plane of a solution of \eref{eq:lagrange_charpit} generates a curve on $\surface$ via the mapping $t\mapsto\rv(u(t),v(t))$; this is  a characteristic line on $\surface$.

Letting $\sigma$ be the arch-length 
parameter along a characteristic line $\curve$ of \eref{eq:lagrange_charpit} on $\surface$ we write
\begin{equation}
	\label{eq:r_dot}
	\dot{\rv}=\dot{u}\rvu+\dot{v}\rvv
\end{equation}
and, by use of \eref{eq:lagrange_charpit}, we easily obtain that
\begin{equation}\label{eq:char_sigma_dot}
\dot\sigma = \|\dot{\rv}\| = r\sqrt{\dot{u}^2 + \dot{v}^2} = \sqrt{B^2 + 1},
\end{equation}
so that the unit tangent vector $\vt$ to $\curve$ is given by
\begin{equation}\label{eq:char_t}
\vt = \frac{\dot{\rv}}{\dot\sigma} =  r\frac{\dot{u}\eu + \dot{v}\ev}{\sqrt{B^2 + 1}}
= 
\frac{(\cos\alpha-B\sin\alpha)\eu + (\sin\alpha+B\cos\alpha)\ev}{\sqrt{B^2 + 1}},
\end{equation}
whence it readily follows that 
\begin{equation}\label{eq:char_t_perp}
\vtp =  \frac{-(\sin\alpha+B\cos\alpha)\eu + (\cos\alpha-B\sin\alpha)\ev}{\sqrt{B^2 + 1}}.
\end{equation}
Equations \eref{eq:char_t} and \eref{eq:char_t_perp}, combined with \eref{eq:n_representation}, show that 
\begin{equation}
	\label{eq:char_t_t_perp}
	\n\cdot\vt=\frac{1}{\sqrt{B^2+1}}\quad\textrm{and}\quad\n\cdot\vtp=-\frac{B}{\sqrt{B^2+1}},
\end{equation}
and so $\n$ can also be written as in \eref{eq:n_1d}, provided that we set
\begin{equation}
	\label{eq:char_gamma}
	\tan\gamma=-B.
\end{equation}

To determine the geodesic curvature of the characteristic line $\curve$, from \eref{eq:char_t} we first compute 
\begin{equation}\label{eq:char_t_prime}
\vt' =\alpha'\vtp + \frac{\cos\alpha-B\sin\alpha}{\sqrt{B^2 + 1}}[(\vt\cdot\vc)\ev + (\vt\cdot\vdu)\vnu] 
+ \frac{\sin\alpha+B\cos\alpha}{\sqrt{B^2 + 1}}[-(\vt\cdot\vc)\eu + (\vt\cdot\vdv)\vnu],
\end{equation}
where a prime $\null'$ denotes differentiation with respect to $\sigma$ and use has also been made of \eref{eq:gliding_laws} and the identities $\eu^\prime=(\grads\eu)\vt$, $\ev^\prime=(\grads\ev)\vt$. Since, by \eref{eq:char_sigma_dot}, 
\begin{equation}
	\label{eq:char_alpha_prime}
	\alpha'=\frac{\dot{\alpha}}{\dot{\sigma}}=\frac{\dot{\alpha}}{\sqrt{B^2+1}},
\end{equation}
by \eref{eq:char_t_perp}, \eref{eq:char_t}, and \eref{eq:lagrange_charpit}, we arrive from \eref{eq:char_t_prime} and \eref{eq:char_alpha_prime} to
\begin{equation}\label{eq:null_kappa}
\kappa_g = \vt'\cdot\vtp = \frac{\dot{\alpha}}{\sqrt{B^2+1}} + \vt\cdot\vc = 0,
\end{equation}
which says that $\curve$ is a geodesic of $\surface$. We shall call $\curve$ a \emph{characteristic geodesic}, for short.

Thus \eref{eq:quasi_uniformity_definition} and \eref{eq:char_gamma} make \eref{eq:gamma_equation} identically satisfied along $\curve$. Also equation \eref{eq:gamma_equation_bis} is identically satisfied with $g=0$, and so $\n$ is parallel transported along $\curve$.

The following statement summarizes our main conclusions.
\begin{thm}\label{thm:geodesics}
On a regular surface $\surf$, the characteristic lines that propagate a quasi-uniform distortion  are geodesics along which the director $\n$ is   parallel transported.
\end{thm}
Theorem \ref{thm:geodesics} suggests that quasi-uniform distortions are the most natural surface distortions that could emerge  as possible nematic ground states, determined solely by the geometry of the surface.

In the following section, we shall further explore the connection between nematic quasi-uniformity and intrinsic geometry of surfaces.

\section{Character evolution}\label{sec:character}
Along with the frame $\framee$ associated with isothermal coordinates of $\surface$, another moving frame can be introduced on $\surface$, that is, $\framen$. The corresponding gliding laws, which replace \eref{eq:gliding_laws}, then read as
\begin{equation}\label{eq:gliding_laws_n}
	\system{
		&\grads\n = \np\otimes\vc^\ast + \vnu\otimes\vd_1, \\
		&\grads\np = -\n\otimes\vc^\ast + \vnu\otimes\vd_2, \\
		&\grads\vnu = -\n\otimes\vd_1 - \np\otimes\vd_2,
	}
\end{equation}
where $(\cv^\ast,\vd_1,\vd_2)$ are the appropriate connectors. In particular, \eref{eq:K} will here be replaced by
\begin{equation}
	\label{eq:K_replaced}
	K=\vd_1\times\vd_2\cdot\normal.
\end{equation}
Combining the first equation in \eref{eq:gliding_laws_n} and \eref{eq:surface_gradient_representation}, we easily identify a general representation for the spin connector,
\begin{equation}
	\label{eq:spin_connector_star}
	\cv^\ast=-b_\perp\n+\splay\np,
\end{equation}
which, by \eref{eq:character_definition}, for a quasi-uniform distortion reads as 
\begin{equation}
	\label{eq:spin_connector_quasi}
	\vc^\ast=f(-B_0\n+S_0\np),
\end{equation}
where both $B_0$ and $S_0$ are constant and $f$ is the quasi-uniform character of the director field $\n$.

We wish to find the equation that governs the evolution of $f$ along a characteristic geodesic that conveys a quasi-uniform field $\n$. This will follow from the integrability condition for $\grads\n$, as expressed by the first equation in \eref{eq:gliding_laws_n} when $\vc^\ast$ is as in \eref{eq:spin_connector_quasi}. As shown in \cite{pedrini:surface} (see also \cite{sonnet:bending-neutral,sonnet:variational}), the necessary and sufficient condition for the integrability of $\grads\n$ is given by
\begin{equation}
	\label{eq:integrability_condition}
	2\skw(\nablastwo\n)=2\skw((\grads\n)\curvature\otimes\normal),
\end{equation}
where the operator $\skw$ acts as follows on a generic \emph{triadic} third-rank tensor $\va_1\otimes\va_2\otimes\va_3$,
\begin{equation}
	\label{eq:skw_action}
	2\skw(\va_1\otimes\va_2\otimes\va_3)=\va_1\otimes\va_2\otimes\va_3-\va_1\otimes\va_3\otimes\va_2.
\end{equation}
Thus equation \eref{eq:integrability_condition} has the general structure
\begin{equation}
	\label{eq:integrability_structure}
	\n\otimes\W_1+\np\otimes\W_2+\normal\otimes\W_3=\bm{0},
\end{equation}
where $\W_1$, $\W_2$, and $\W_3$ are skew-symmetric second-rank tensors. Since $\framen$ is a basis, \eref{eq:integrability_structure} amounts to require that
\begin{equation}
	\label{eq:w_1_2_3}
	\wv_1=\wv_2=\wv_3=\bm{0},
\end{equation}
where $\wv_1$, $\wv_2$, and $\wv_3$ are the axial vectors of $\W_1$, $\W_2$, and $\W_3$, respectively.
For our purposes it will suffice to determine $\wv_2$.

It follows from \eref{eq:gliding_laws_n} and \eref{eq:spin_connector_quasi} that
\begin{equation}
	\label{eq:integrability_1}
	\eqalign{
	\np\cdot\nablastwo\n&=f^2[S_0B_0(\n\otimes\n-\np\otimes\np)+B_0^2\np\otimes\n-S_0^2\n\otimes\np]\\
	&+f(-B_0\normal\otimes\vd_1+S_0\normal\otimes\vd_2)
	-B_0\n\otimes\grads f+S_0\np\otimes\grads f-\vd_1\otimes\vd_2
}
\end{equation}
and
\begin{equation}
	\label{eq:integrability_2}
	\np\cdot(\grads\n)\curvature\otimes\normal=f(-S_0\vd_2\otimes\normal+B_0\vd_1\otimes\normal).
\end{equation}
With little labour we extract $\wv_2$ from \eref{eq:integrability_1} and \eref{eq:integrability_2},
\begin{equation}
	\label{eq:w_2}
	\wv_2=[(S_0^2+B_0^2)f^2+(S_0\n+B_0\np)\cdot\grads f]\normal+\vd_1\times\vd_2.
\end{equation}
Thus, also by \eref{eq:K_replaced}, \eref{eq:w_1_2_3} implies that
\begin{equation}
	\label{eq:f_pre_equation}
	(S_0^2+B_0^2)f^2+(S_0\n+B_0\np)\cdot\grads f+K=0.
\end{equation}
Finally, by \eref{eq:char_t_t_perp}, the unit tangent $\vt$ to a characteristic geodetic can also be written as
\begin{eqnarray}
	\vt&=&\frac{1}{\sqrt{B^2+1}}(\n+B\np)\label{eq:t_rewritten_1}\\
	&=&\frac{\sgn(S_0)}{\sqrt{S_0^2+B_0^2}}(S_0\n+B_0\np),\label{eq:t_rewritten_2}
\end{eqnarray}
where \eref{eq:B_formula} has also been used and $\sgn$ denotes the sign function. By inserting \eref{eq:t_rewritten_2} in \eref{eq:f_pre_equation} we obtained the desired result,
\begin{equation}
	\label{eq:f_formula}
	(S_0^2+B_0^2)f^2+\sgn(S_0)\sqrt{S_0^2+B_0^2}f'+K=0,
\end{equation}
where $f'=\grads f\cdot\vt$ is the derivative with respect to the arc-length $\sigma$ of $\curve$.

Since the sign of $S_0$ is arbitrary, use of \eref{eq:scaling_identity} reduces \eref{eq:f_formula} to a simpler Riccati equation,
\begin{equation}
	\label{eq:character_evolution}
	f^2\pm f'+K=0,
\end{equation}
which determines $f$ along $\curve$, to within a sign and an arbitrary  constant (which can be fixed by prescribing $f$ at one point of $\curve$).

As already remarked, a uniform distortion on $\surface$ is a special quasi-uniform distortion for which $f^2\equiv1$. Then equation \eref{eq:character_evolution} reduces to a known condition, proved in \cite{niv:geometric},
\begin{equation}
	\label{eq:character_uniformity}
	K\equiv-1,
\end{equation}
which requires $\surface$ to be a \emph{pseudospherical} surface, according to the definition of Beltrami \cite[p.\,22]{needham:visual} (see also \cite{pedrini:surface}). Comparing \eref{eq:character_evolution} and \eref{eq:character_uniformity} also shows that a quasi-uniform distortion can live on a surface $\surface$ with generic curvature: the quasi-uniformity character will simply adapt to its intrinsic geometry.

In Appendix~\ref{sec:revolution}, equation \eref{eq:character_evolution} is solved explicitly for the special surfaces studied in the following section.

\section{Applications}\label{sec:applications}
In this section, we apply the theory developed above to the explicit construction of quasi-uniform nematic fields on two exemplary surfaces, different in both shape and extension, namely, a sphere and a catenoid. We shall be concerned both with  fields that can be extended globally (away from point defects mandated by topological requirements) and with fields prescribed on a curve, which may or may not be extended globally.

\subsection{Global quasi-uniformity on a sphere}\label{sec:sphere}
We first consider the two-dimensional unit sphere $\sphere$, with normal vector $\normal=\er$. 
\begin{figure}
	\centering
	\includegraphics[width=.3\textwidth]{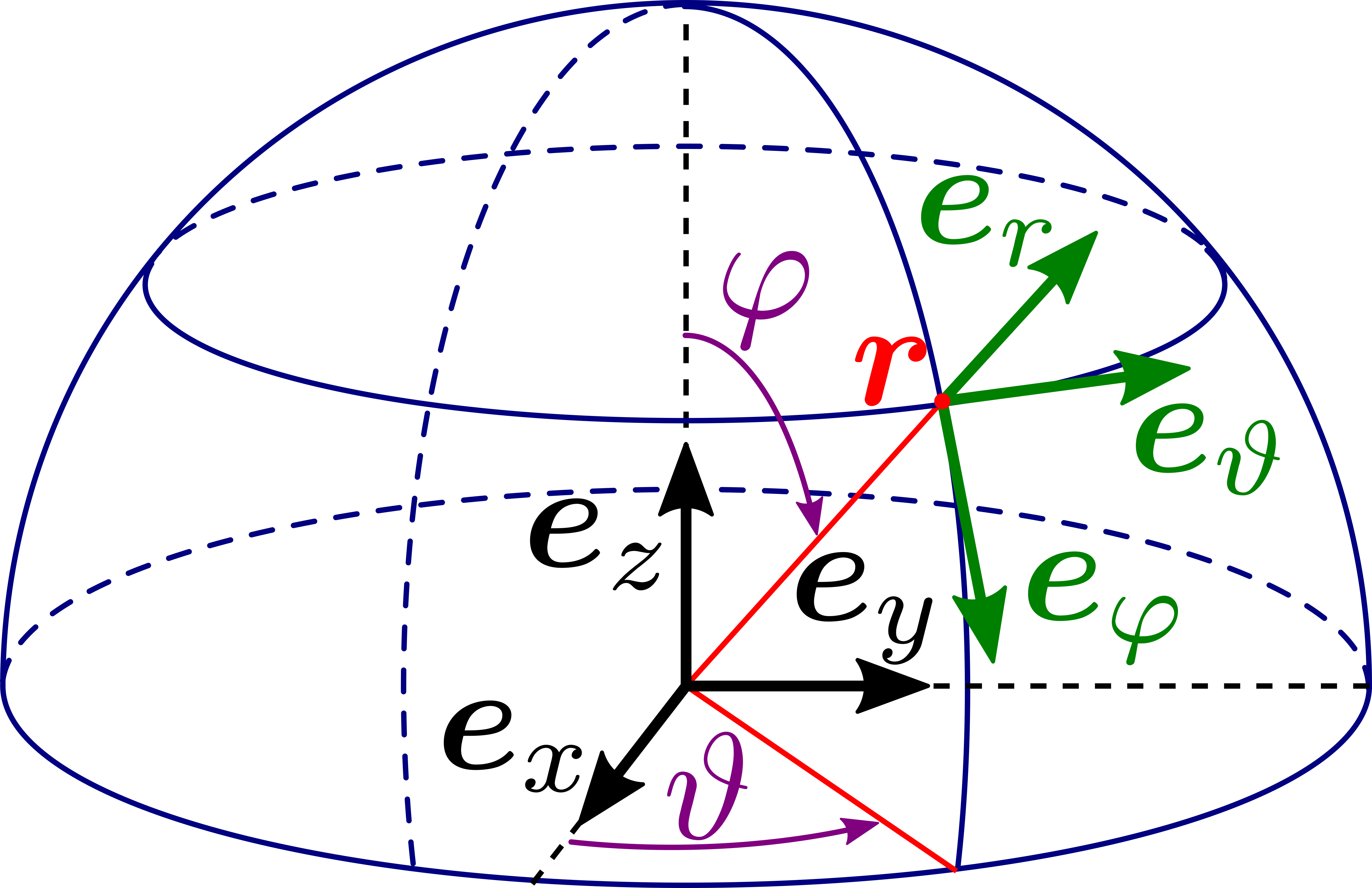}
	\caption{The moving frame $(\ep,\et,\normal)$ associated with standard spherical coordinates.}
	\label{fig:polar}
\end{figure}
Figure~\ref{fig:polar} shows the moving frame $(\ep,\et,\normal)$ associated with standard spherical coordinates $(\varphi,\vartheta)$ on $\sphere$, where $\varphi{\in[0,\pi]}$ is the \emph{polar} angle and $\vartheta\in[0,2\pi)$ is the \emph{azimuthal} angle; the frame $(\ep,\et,\normal)$ is related as follows to a Cartesian frame $\frameC$,
\begin{equation}\label{eq:spherical}
	\system{
		\ep &= \cos\varphi\cos\vartheta\ex + \cos\varphi\sin\vartheta\ey - \sin\varphi\ez, \\
		\et &= - \sin\vartheta\ex + \cos\vartheta\ey,\\
		\er &= \sin\varphi\cos\vartheta\ex + \sin\varphi\sin\vartheta\ey + \cos\varphi\ez.
		}
\end{equation}
Standard computations then show that 
\begin{equation}\label{eq:spherical_grads}
	\system{
		\grads\ep &:= -\er\otimes\ep + \cot\varphi\et\otimes\et, \\
		\grads\et &:= -\er\otimes\et - \cot\varphi\ep\otimes\et,\\
		\grads\er &:= \ep\otimes\ep + \et\otimes\et.
		}
\end{equation}

Spherical coordinates fail to be isothermal; to make them so, we set
\begin{equation}\label{eq:sphere_iso}
u:=-\ln\frac{1+\cos\varphi}{\sin\varphi}
\quad\textrm{and}\quad
v:=\vartheta.
\end{equation}
It readily follows from \eref{eq:sphere_iso} that
\begin{equation}
	\label{eq:u_derivatives}
	u_{,\varphi}=\frac1{\sin\varphi}\quad\textrm{and}\quad v_{,\vartheta}=1.
\end{equation}
With this choice,
\begin{equation}\label{eq:r_sphere}
\rv(u,v) = \er,
\quad
\rvu = \sin\varphi\ep,
\quad
\rvv = \sin\varphi\et,
\quad
r=\sin\varphi,
\end{equation}
so that the moving frame $\framee$ can be identified with $(\ep,\et,\normal)$ and, by \eref{eq:spherical_grads}, its connectors are given by
\begin{equation}\label{eq:sphere_connectors}
\vc = \cot\varphi\et, 
\qquad\vdu = -\ep, 
\qquad\vdv = -\et.
\end{equation}

We represent the director field $\n$ as
\begin{equation}\label{eq:field}
\n = \cos\alpha\ep + \sin\alpha\et,
\end{equation}
where $\alpha = \alpha(\varphi,\vartheta)$.
Since, by \eref{eq:u_derivatives}, $\au=\alphap\sin\varphi$ and $\av=\alphat$, use of \eref{eq:sphere_connectors} gives equation \eref{eq:quasi_pde} the following form,
\begin{equation}\label{eq:sphere_pde}
\alphap\sin\varphi(\cos\alpha-B\sin\alpha) + 
\alphat(\sin\alpha+B\cos\alpha) 
= - (\sin\alpha+B\cos\alpha)\cos\varphi.
\end{equation}
We seek a \emph{global} solution of \eref{eq:sphere_pde} on the unit sphere. 

According to Theorem~\ref{thm:geodesics}, the characteristic lines of \eref{eq:sphere_pde} are the \emph{great circles} of $\sphere$, along which, by \eref{eq:char_gamma}, $\n$ makes a constant angle with the tangent. Since thus $\n$ exhibits a defect whenever two such characteristic geodesics meet, the only global solution of \eref{eq:sphere_pde} is obtained when all characteristic geodesics are the meridians passing for the selected poles (where $\sin\varphi=0$), in accordance with the theorem of Poincar\'e \cite{poincare:courbes} that prescribes the total topological charge $\chi$ of defects of $\n$ on $\sphere$ to be $\chi=+2$. Orienting the meridians of $\sphere$ with $\vt=\ep$, we set
\begin{equation}
\alpha(\varphi,\vartheta) \equiv  -\arctan B
\end{equation}
and from \eref{eq:field} we arrive at the following representation for $\n$,
\begin{equation}\label{eq:sphere_n_global}
\n = \frac1{\sqrt{B^2+1}}(\ep - B\et),
\end{equation}
whose field lines are \emph{loxodromes}, which cut all meridians at the same angle $\gamma=-\arctan B$. 

In summary, by \eref{eq:spherical_grads}, the director field \eref{eq:sphere_n_global} is the only 
quasi-uniform distortion with (constant) anisotropy $B$ covering an entire sphere, In spherical coordinates $(\varphi,\vartheta)$, its distortion characteristics are given by
\begin{equation}\label{eq:representation_formulae}
\splay = \frac{\cot\varphi}{\sqrt{B^2+1}}
\quad
\textrm{and}
\quad
\bp = \frac{B\cot\varphi}{\sqrt{B^2+1}}.
\end{equation}

\begin{figure}[h!]
\centering
\begin{subfigure}[b]{0.3\textwidth}
\centering
\includegraphics[width=\textwidth]{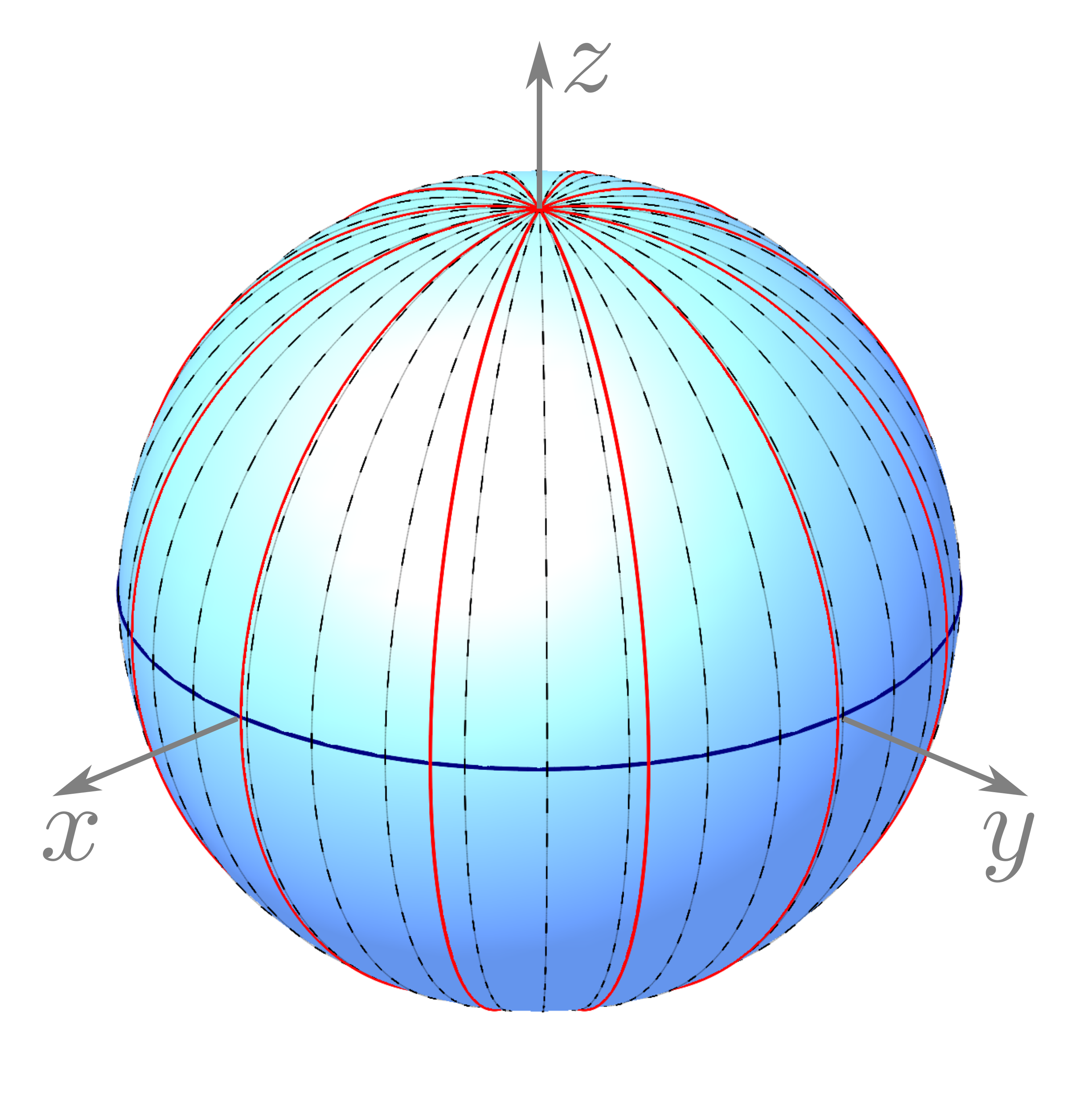}
\caption{$B = 0$}
\end{subfigure}
$\quad$
\begin{subfigure}[b]{0.3\textwidth}
\centering
\includegraphics[width=\textwidth]{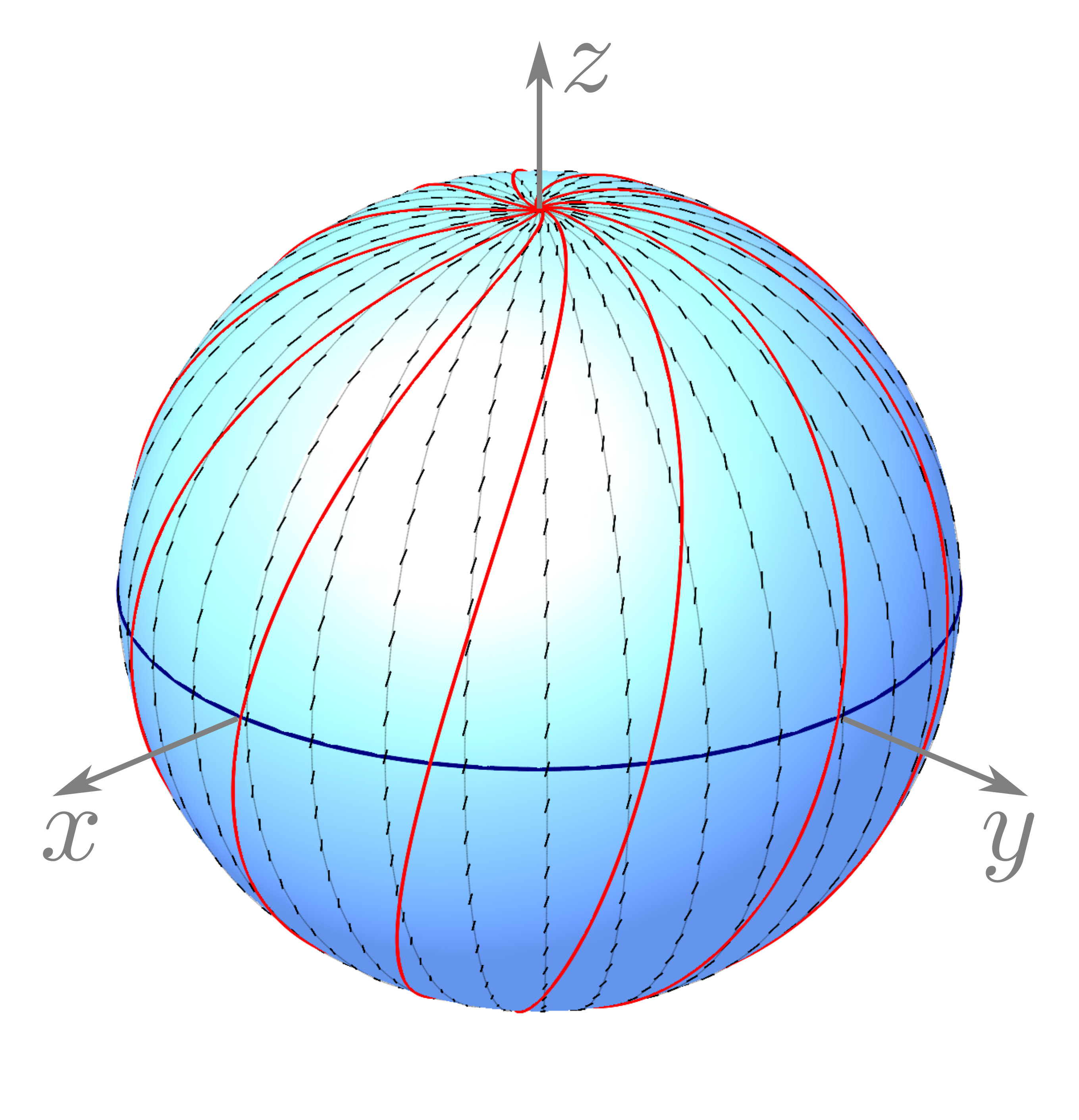}
\caption{$B = \frac14$}
\end{subfigure}
$\quad$
\begin{subfigure}[b]{0.3\textwidth}
\centering
\includegraphics[width=\textwidth]{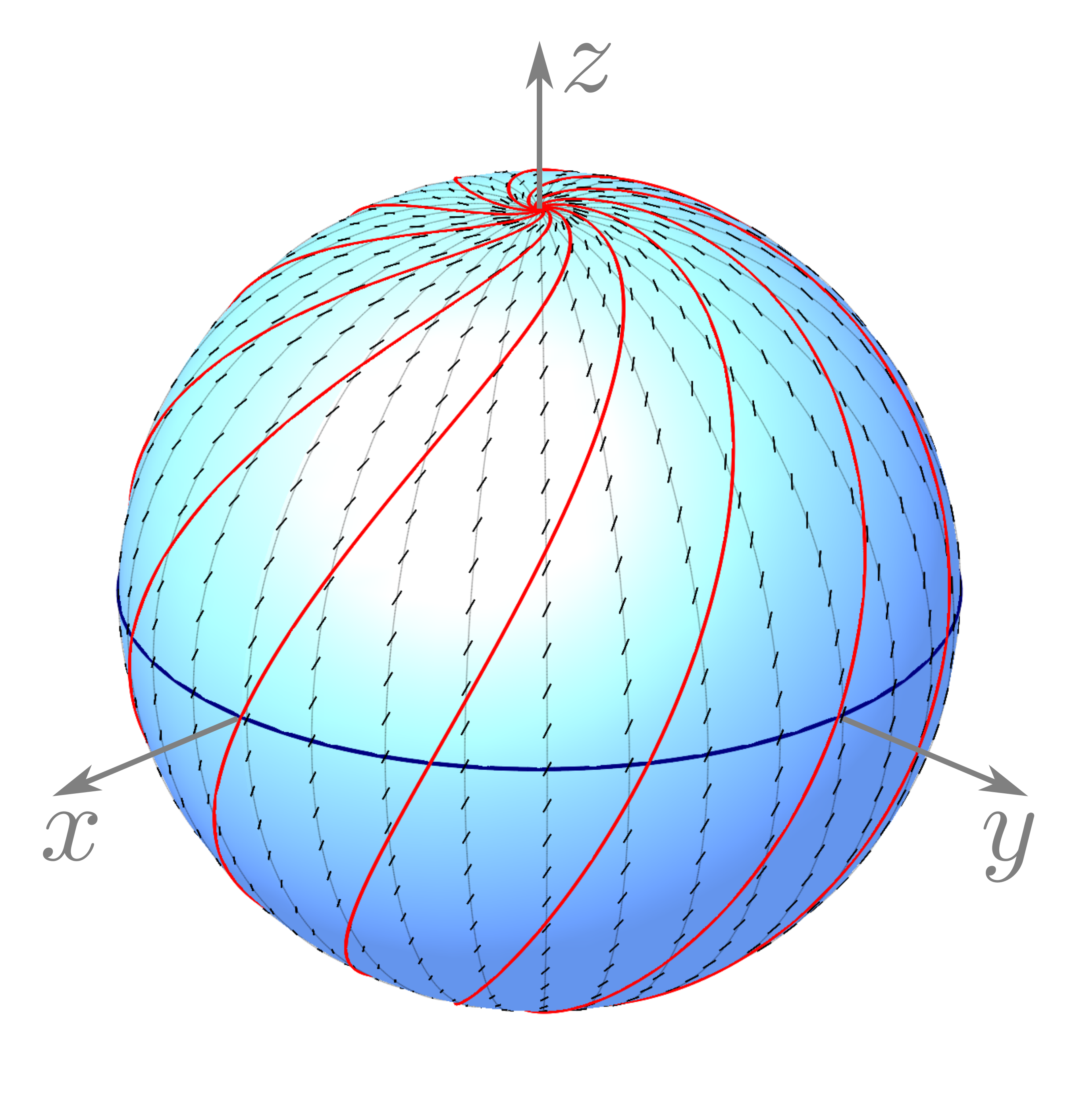}
\caption{$B = \frac12$}
\end{subfigure}
\\
\begin{subfigure}[b]{0.3\textwidth}
\centering
\includegraphics[width=\textwidth]{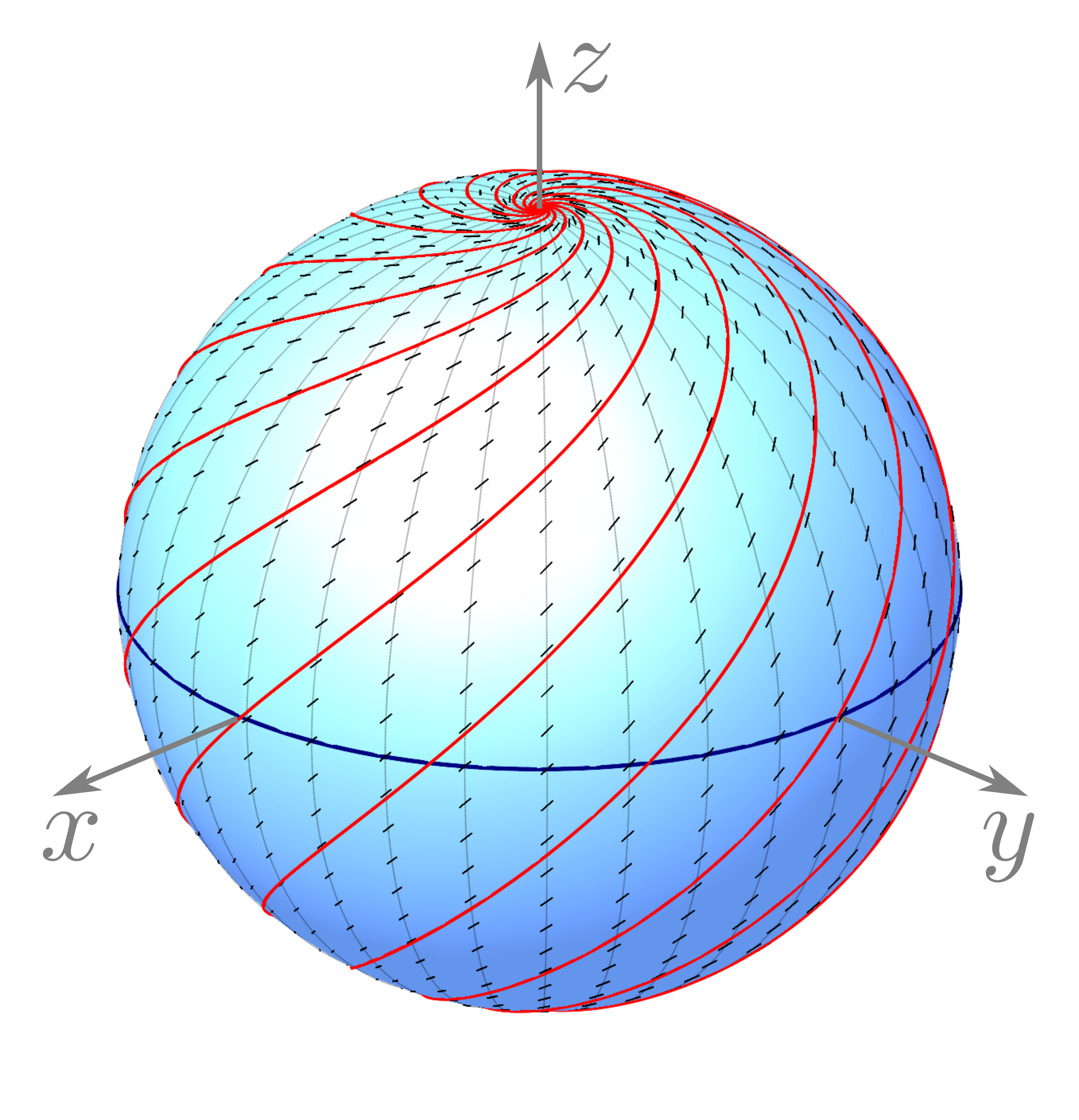}
\caption{$B = 1$}\label{fig:lox1}
\end{subfigure}
$\quad$
\begin{subfigure}[b]{0.3\textwidth}
\centering
\includegraphics[width=\textwidth]{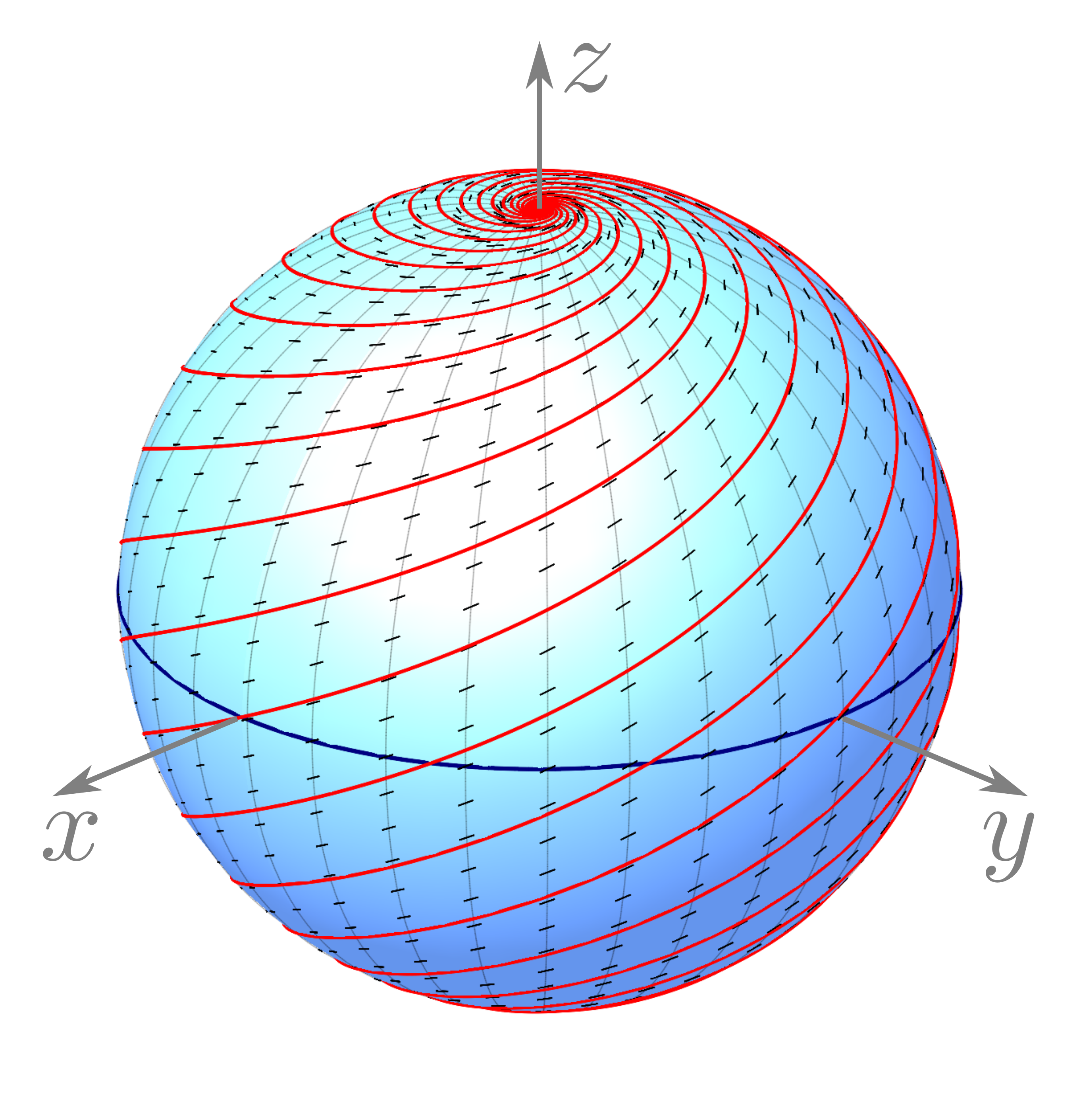}
\caption{$B = 2$}
\end{subfigure}
$\quad$
\begin{subfigure}[b]{0.3\textwidth}
\centering
\includegraphics[width=\textwidth]{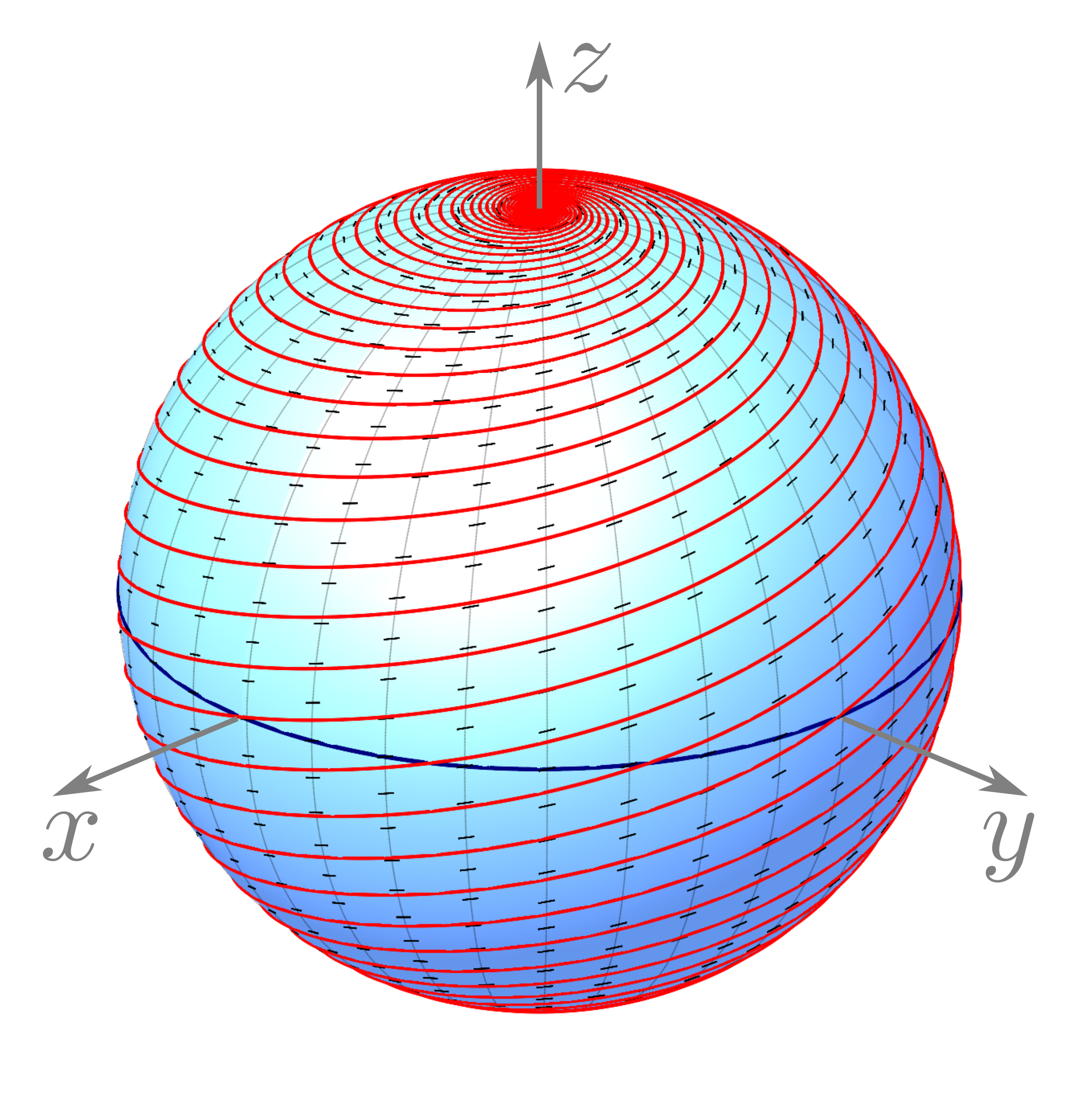}
\caption{$B = 4$}
\end{subfigure}
\caption{Examples of global quasi-uniform distortions $\n$ on a sphere as described by \eref{eq:sphere_n_global} for different values of the (constant) distortion anisotropy $B$. Directors are represented as (black) headless vectors, whose field lines (in red) are loxodromes of the meridians (gray lines). The equator is marked in blue. A pure-splay distortion occurs when $B=0$, while a pure-bend distortion is approached as $B\to\infty$.}
\label{fig:alpha_constant}
\end{figure}
Figure~\ref{fig:alpha_constant} shows some examples of global quai-uniform distortions in \eref{eq:sphere_n_global}, for different values of $B>0$. A \emph{pure-splay} or a \emph{pure-bend} distortion occurs when the field lines of $\n$ are either the meridians or the parallels of the sphere, which is the case in  \eref{eq:sphere_n_global} for either $B=0$ or $B\to\infty$, respectively.

Since quation \eref{eq:sphere_pde} is invariant under the transformation
\begin{equation}\label{eq:S} 
B\mapsto-B,\qquad  \varphi\mapsto\pi-\varphi,\qquad  \vartheta\mapsto \vartheta,\qquad  \alpha\mapsto -\alpha,
\end{equation}
a change in sign of $B$ simply produces a change in the sense of winding of $\n$: the corresponding pictures are produced by mirroring through the equator the images in Figure~\ref{fig:alpha_constant}. 

In Appendix~\ref{app:transports}, we confirm by direct inspection that for the field $\n$ in \eref{eq:sphere_n_global} there are no other geodesics of the sphere along which $\n$ is parallel transported, apart from the meridians and the equator.

\subsection{Local quasi-uniformity on a sphere}
The equator $\curve_0$ parallel transports all global quasi-uniform distortions on the sphere. One might then ask whether a (non-global) quasi-uniform distortion  could arise by prescribing $\n$ to be uniformly (instead of parallel) transported along $\curve_0$. Here, we address this question, demonstrating the remarkable diversity of quasi-uniform distortions that can be generated locally on a surface.

Let
\begin{equation}
	\label{eq:alpha_0_definition}
	\alpha_0(\vartheta):=\alpha\left(\frac{\pi}{2},\vartheta\right)
\end{equation}
be the trace of the angle $\alpha$ in \eref{eq:field} on the equator $\curve_0$. Demanding that $\n$ be uniform on $\curve_0$ requires $\alpha_0'$ to be constant. To ensure continuity of $\n$ on $\curve_0$ (to within the nematic symmetry), we set
\begin{equation}\label{eq:alpha_0_example}
\alpha_0(\vartheta) = (m-1)\vartheta + c_0,
\end{equation}
where $m$ is a \emph{half}-integer and $c_0\in\R$. For $m\neq1$, the constant $c_0$ in \eref{eq:alpha_0_example} is inessential and can be chosen arbitrarily, as changing its value  would only produce an overall rotation of the field $\n$ around the polar axis of $\sphere$ (the $z$-axis in Figure~\ref{fig:polar}). For $m=1$, instead, changing $c_0$  describes all (different) distortions whose field lines are loxodromes on the sphere.

Extending locally around $\curve_0$ a prescribed field $\n$ to a quasi-uniform distortion with a prescribed $B$ reduces to solving equation \eref{eq:sphere_pde}
subject to the Cauchy data in \eref{eq:alpha_0_example}. As learned in Section~\ref{sec:geodesic}, 
the characteristic lines of \eref{eq:sphere_pde} are great circles conveying $\n$ at the (constant) angle $\gamma=-\arctan B$. By \eref{eq:t_rewritten_1},  each individual characteristic great circle cuts the equator at an angle $\beta$ such that
\begin{equation}\label{eq:beta_zero}
\tan\beta = \frac{\vt\cdot\ep}{\vt\cdot\et}
= \frac{\cos\alpha_0 - B\sin\alpha_0}{\sin\alpha_0 + B\cos\alpha_0},
\end{equation}
where \eref{eq:field} has also been used.

We now consider separately two special cases of \eref{eq:alpha_0_example}.

\subsubsection{Case $m=1$.}
By \eref{eq:beta_zero}, when
$\alpha_0\equiv c_0\neq-\arctan B$, all characteristic great circles share the same value of $\beta$, thus only differing from one another by a rotation about the polar axis. Each propagates $\n$ with constant angle  away from the equator up to its summit, reached at $\varphi=\varphi^\ast$, after having traversed the length  $\frac\pi2$. The construction of the field $\n$ cannot progress any farther without generating a plethora of defects, produced by the intersection of characteristic great circles as they reverse course beyond their  summit upon reapproaching the equator $\curve_0$.\footnote{This picture applies to the northern hemisphere; for the southern hemisphere, it should be replaced by its mirror image thorough the equator.}

Thus, the quasi-uniform extension around the equator is confined within a spherical zone, bounded by the parallel tangent to the  summits of all  characteristic great circles and its mirror image through the equator. There, $\vt=\et$, and so, by combining \eref{eq:t_rewritten_1} and \eref{eq:beta_zero}, we have that 
\begin{equation}
	\label{eq:alpha_summit}
	\tan\alpha(\varphi^\ast,\vartheta)=\frac{1}{B}.
\end{equation}
Since, by \eref{eq:sphere_connectors}, $c_u=0$ and $c_v=\cot\varphi$, equations \eref{eq:lagrange_charpit}, \eref{eq:u_derivatives}, and \eref{eq:sphere_pde} imply that 
\begin{equation}
\frac{\dd\alpha}{\dd\varphi}=-\frac{\sin\alpha+B\cos\alpha}{\cos\alpha-B\sin\alpha}\cot\varphi,
\end{equation}
which can be integrated at once by separation of variables,
\begin{equation}\label{eq:pre_zone}
\sin\varphi = \left|\frac{\sin\alpha_0+B\cos\alpha_0}{\sin\alpha+B\cos\alpha}\right|.
\end{equation}
It readily follows from \eref{eq:pre_zone} and \eref{eq:alpha_summit} that the  amplitude of the maximal spherical zone where $\n$ can be extended quasi-uniformly is determined by
\begin{equation}\label{eq:sin_phi}
\sin\varphi^* = \frac{|\sin\alpha_0 + B\cos\alpha_0|}{\sqrt{B^2+1}}.
\end{equation}

If we choose $\alpha_0=-\arctan B$, the quasi-uniform extension covers the whole sphere, in accord with \eref{eq:sin_phi}, and the case under study is subsumed under that studied in Section~\ref{sec:sphere} above. We say that this is the \emph{resonant} case. On the contrary, if we choose $\alpha_0=\arctan\frac{1}{B}$, \eref{eq:sin_phi} implies that the maximal zone of extension degenerates into the equator, and no extension actually exists.

\begin{figure}[h!]
\centering
\begin{subfigure}[b]{0.3\textwidth}
\centering
\includegraphics[width=\textwidth]{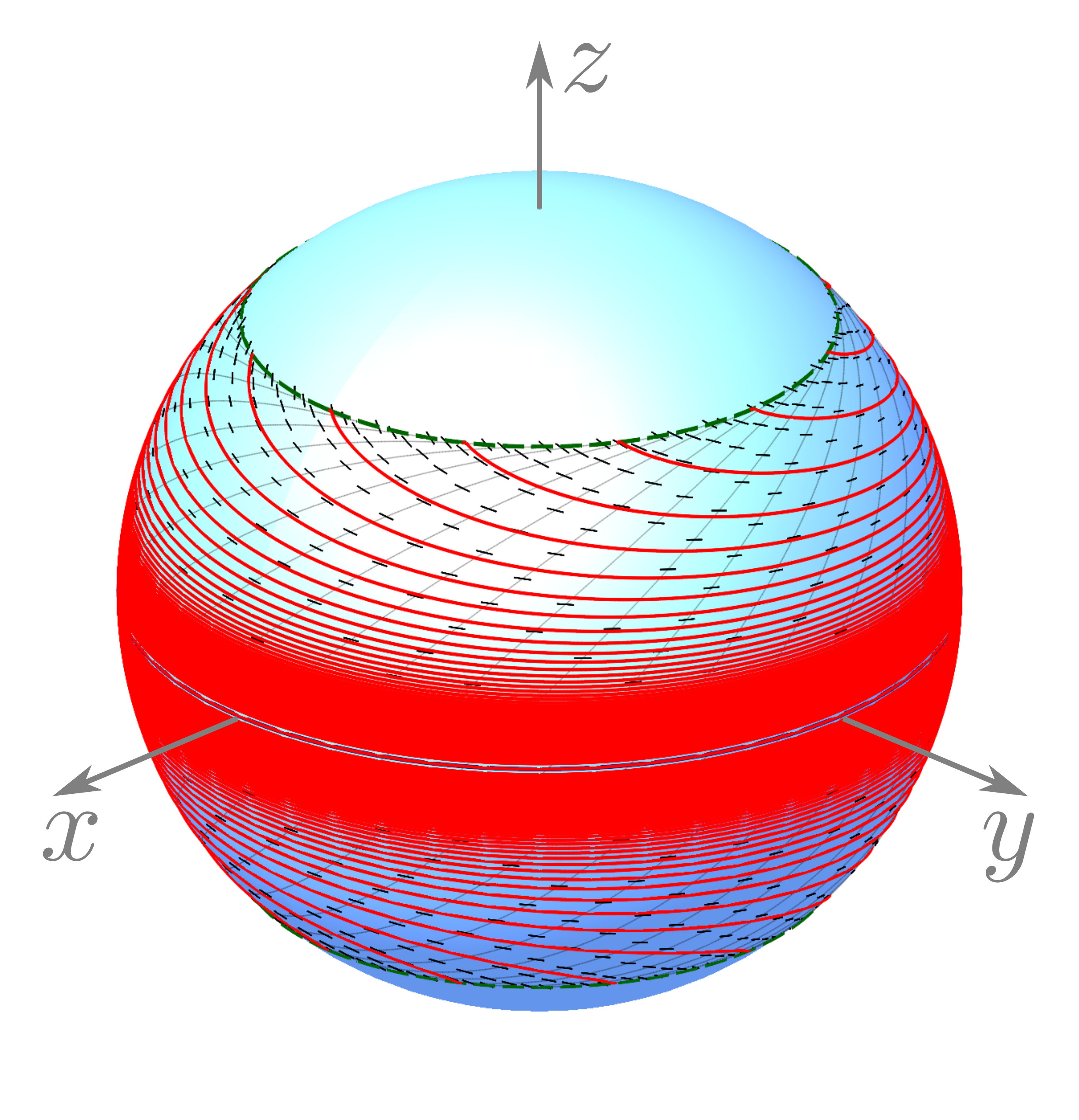}
\caption{$\alpha_0=\frac\pi2$}
\end{subfigure}
\quad
\begin{subfigure}[b]{0.3\textwidth}
\centering
\includegraphics[width=\textwidth]{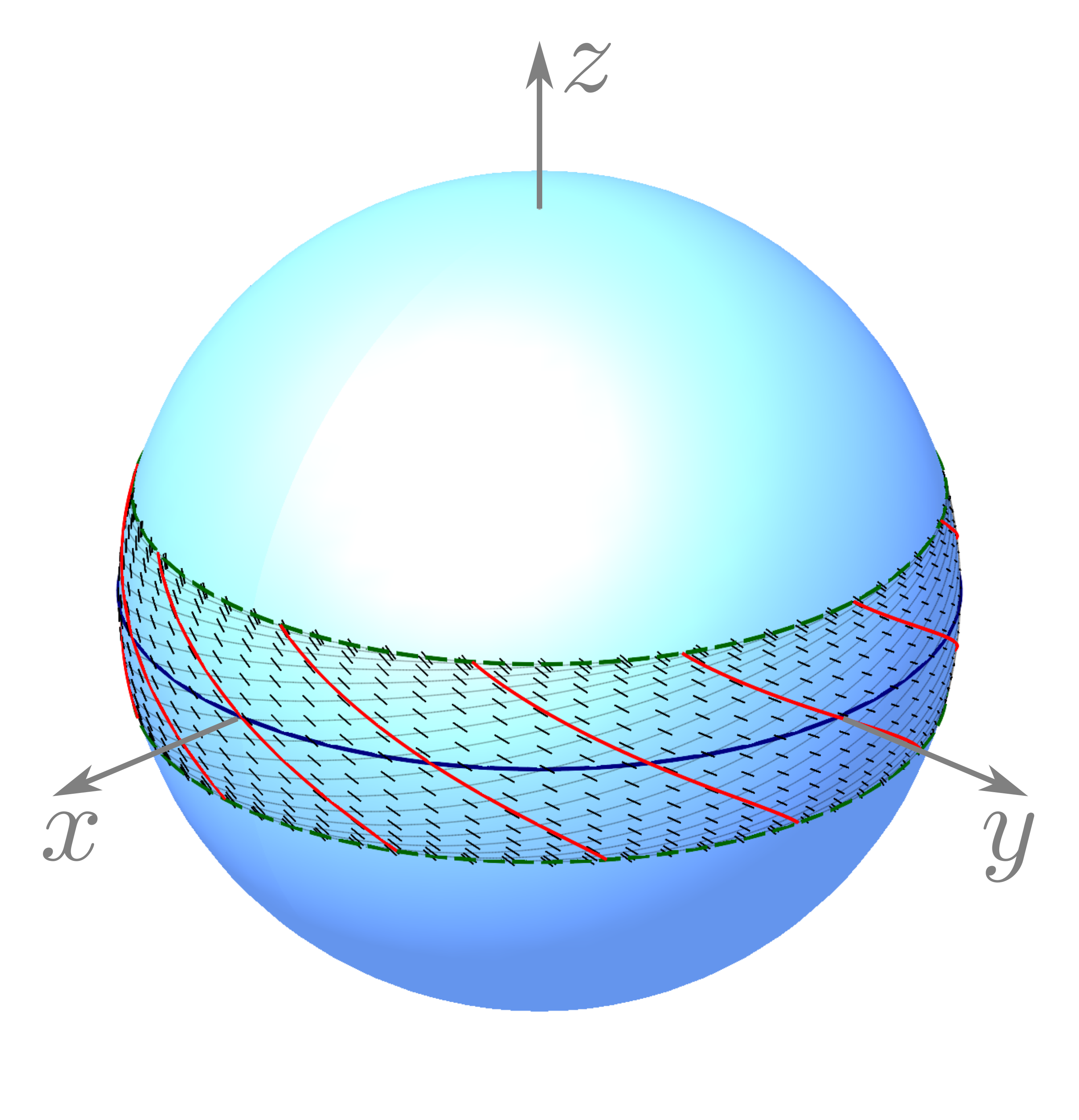}
\caption{$\alpha_0=\frac\pi3$}
\end{subfigure}
\quad
\begin{subfigure}[b]{0.3\textwidth}
\centering
\includegraphics[width=\textwidth]{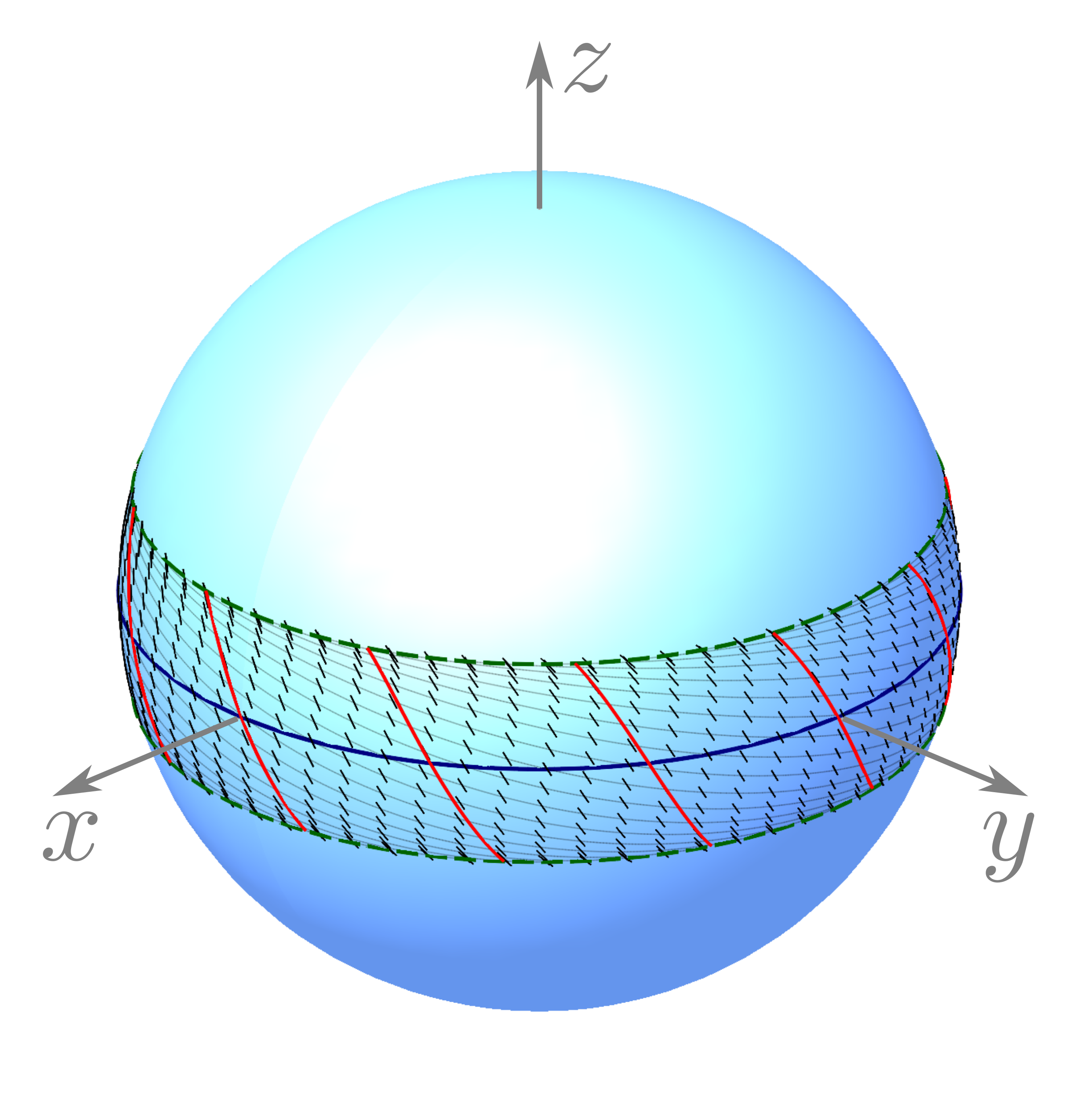}
\caption{$\alpha_0=\frac\pi6$}
\end{subfigure}
\\
\begin{subfigure}[b]{0.3\textwidth}
\centering
\includegraphics[width=\textwidth]{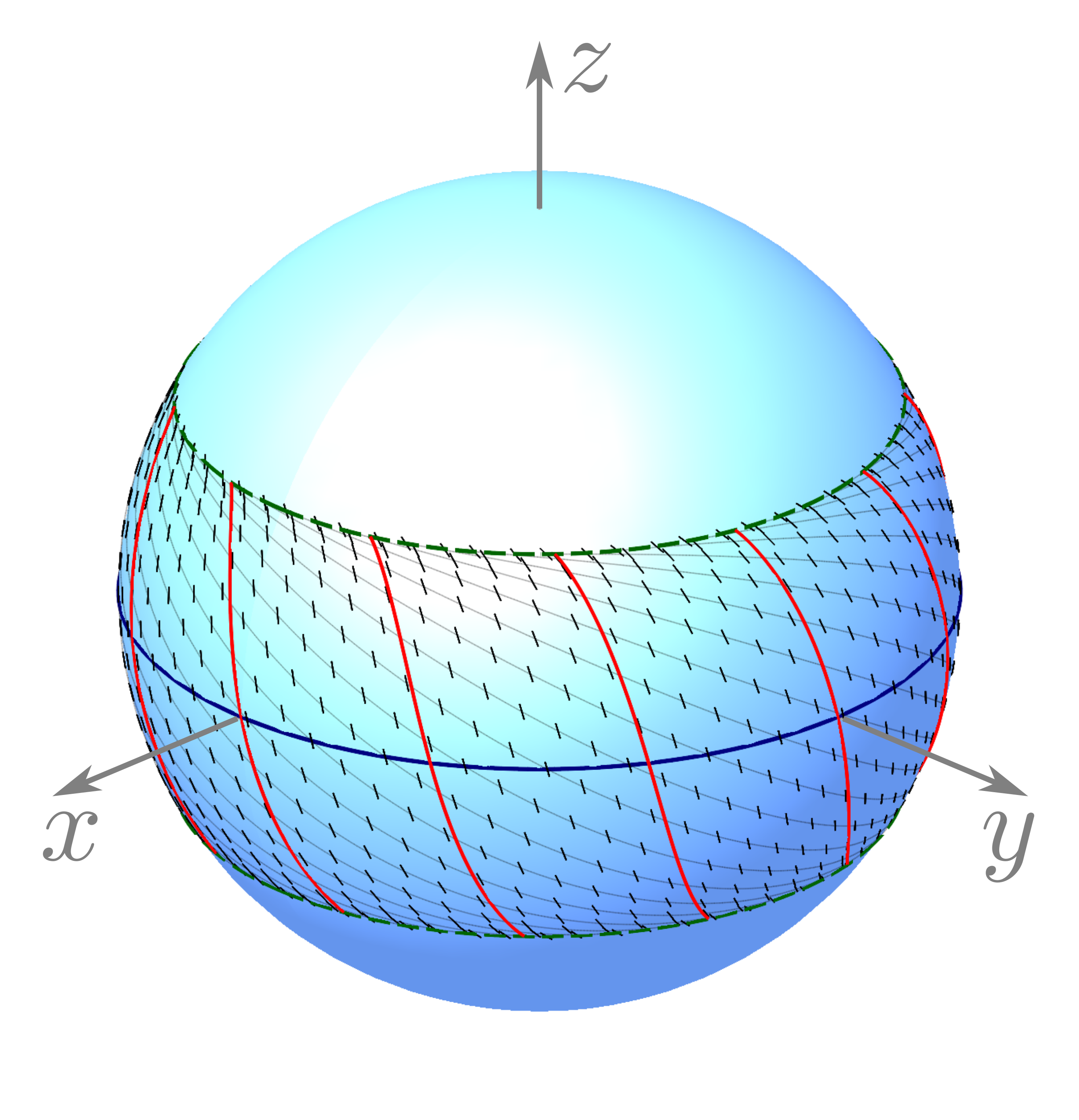}
\caption{$\alpha_0=\frac\pi{12}$}
\end{subfigure}
\quad
\begin{subfigure}[b]{0.3\textwidth}
\centering
\includegraphics[width=\textwidth]{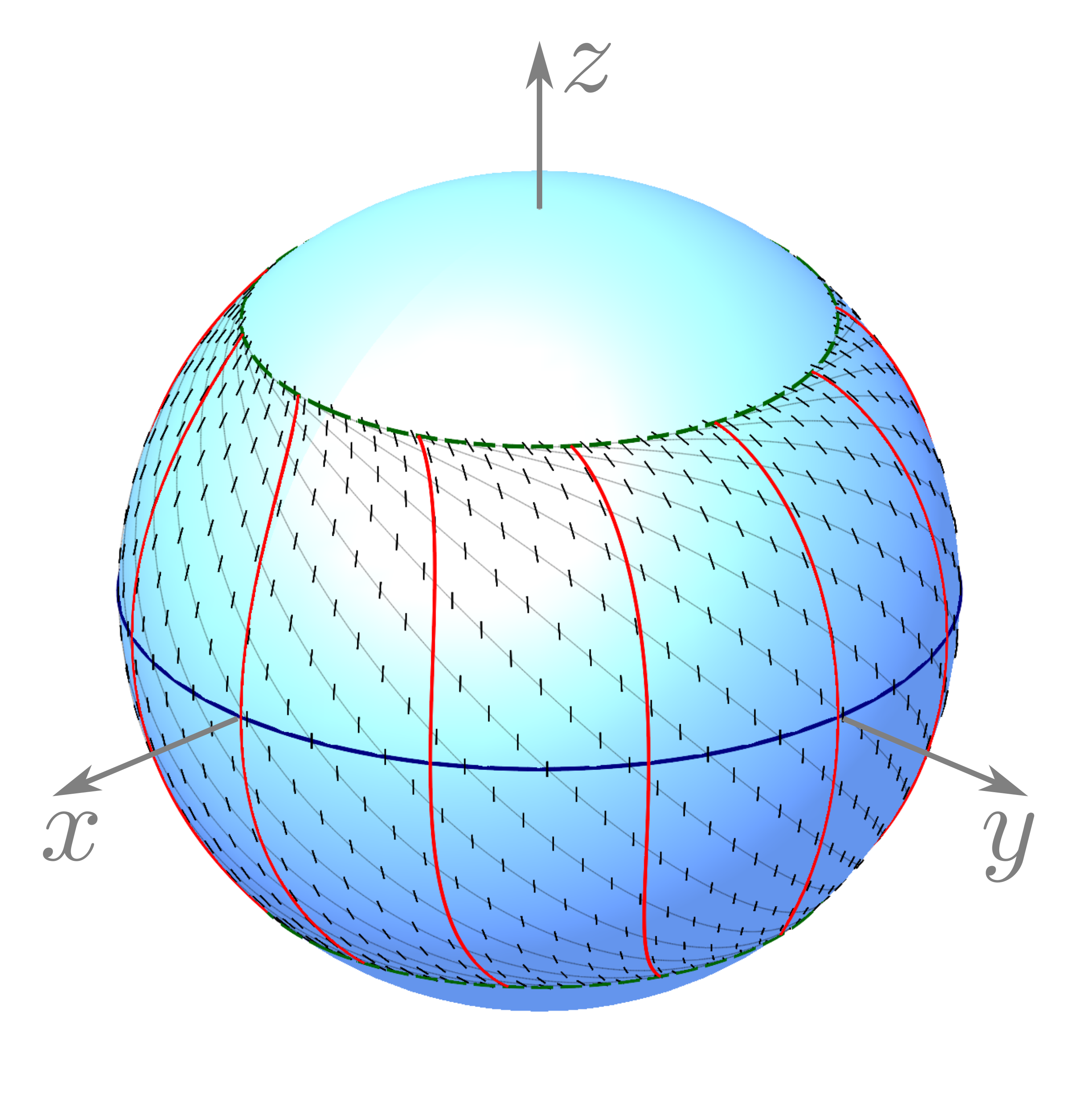}
\caption{$\alpha_0=0$}
\end{subfigure}
\quad
\begin{subfigure}[b]{0.3\textwidth}
\centering
\includegraphics[width=\textwidth]{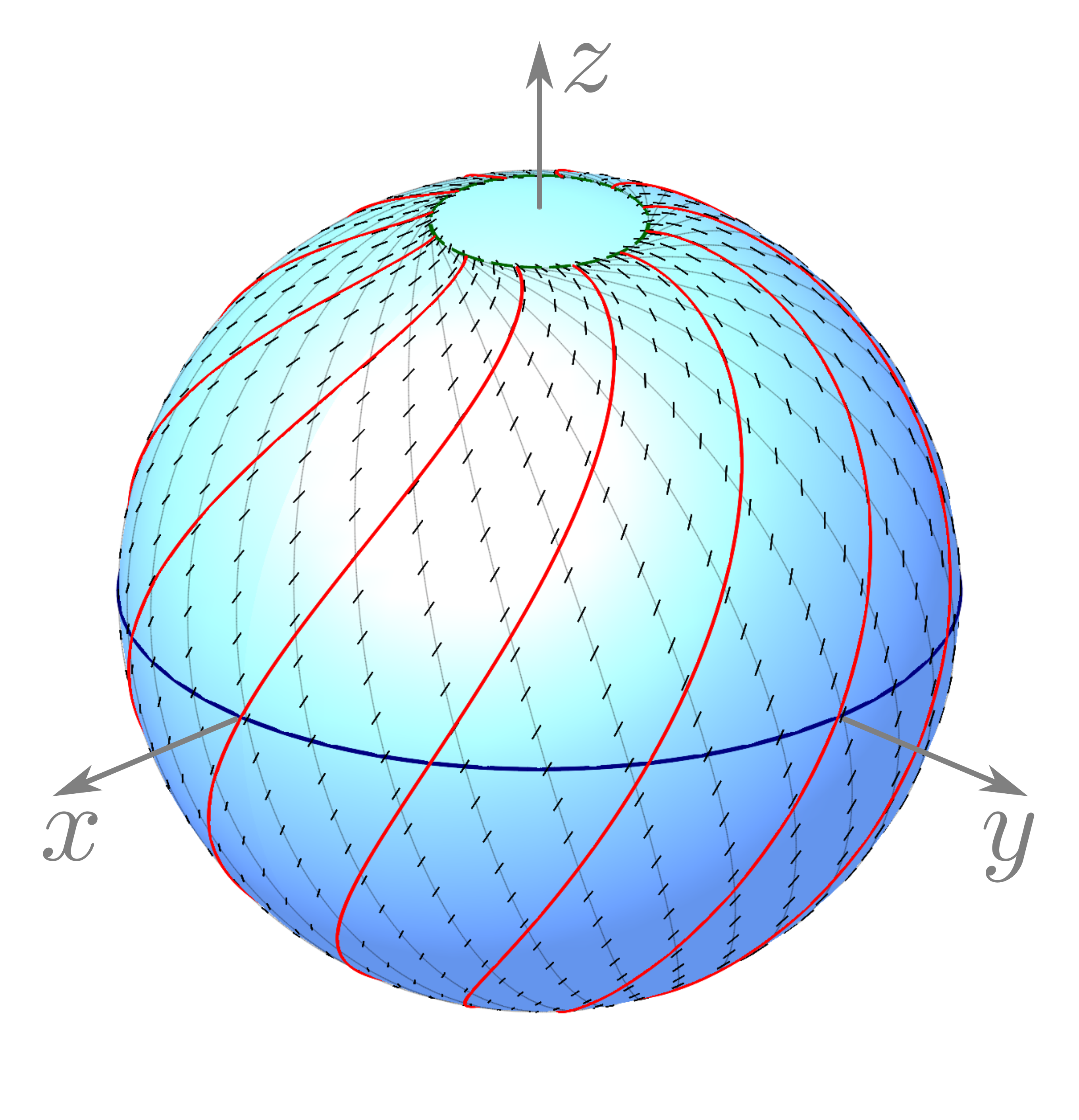}
\caption{$\alpha_0=-\frac\pi6$}
\end{subfigure}
\caption{Field lines (in red) of local quasi-uniform distortions $\n$ with constant angle $\alpha_0$ on the equator (blue line) of the unit sphere, for $B=1$.  Directors are depicted as headless vectors and the parallels delimiting the maximal zone of quasi-uniform extension according to \eref{eq:sin_phi} are dashed green. The characteristic great circles transporting the Cauchy data in \eref{eq:alpha_0_example} are gray. For $\alpha_0=\frac\pi4$, the maximal zone of extension would degenerate into the equator itself.}
\label{fig:alpha_meridians}
\end{figure}

Figure~\ref{fig:alpha_meridians} represents a number of exemplary quasi-uniform extensions for $B=1$ of constant Cauchy data in \eref{eq:alpha_0_example} such that $\alpha_0\neq\pm\frac{\pi}{4}$; the maximal zones of extension are determined according to \eref{eq:sin_phi}.

\subsubsection{Case $m\neq1$.}
Since in this case $\alpha_0$ undergoes a total change of  $2|m-1|\pi$ around the whole equator, there are precisely $2|m-1|$ points $p_i$ on $\curve_0$, identified by as many values of $\vartheta$, designated as $\vartheta_i$, where $\tan\alpha_0=\frac1B$. According to \eref{eq:beta_zero}, the corresponding characteristic great circles emanating from each $p_i$ is the equator itself, which intersects infinitely many other characteristic great circles in any neighbour of $\vartheta_i$. Thus, the boundary condition $\alpha_0$ cannot be extended quasi-uniformly  around $p_i$. 

\begin{figure}[h!]
\centering
\begin{subfigure}[b]{0.3\textwidth}
\centering
\includegraphics[width=\textwidth]{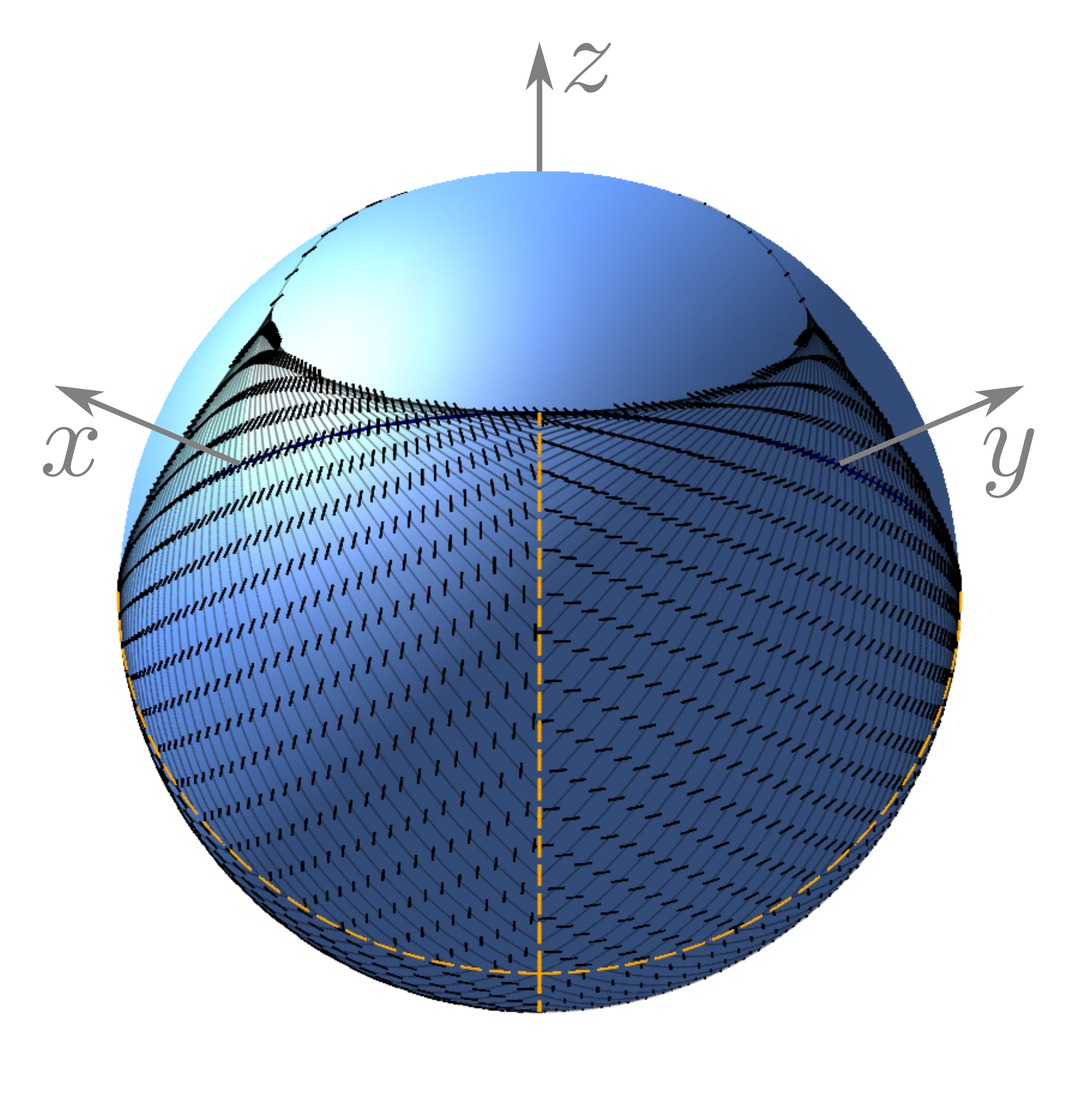}
\caption{$m=3$}
\end{subfigure}
\quad
\begin{subfigure}[b]{0.3\textwidth}
\centering
\includegraphics[width=\textwidth]{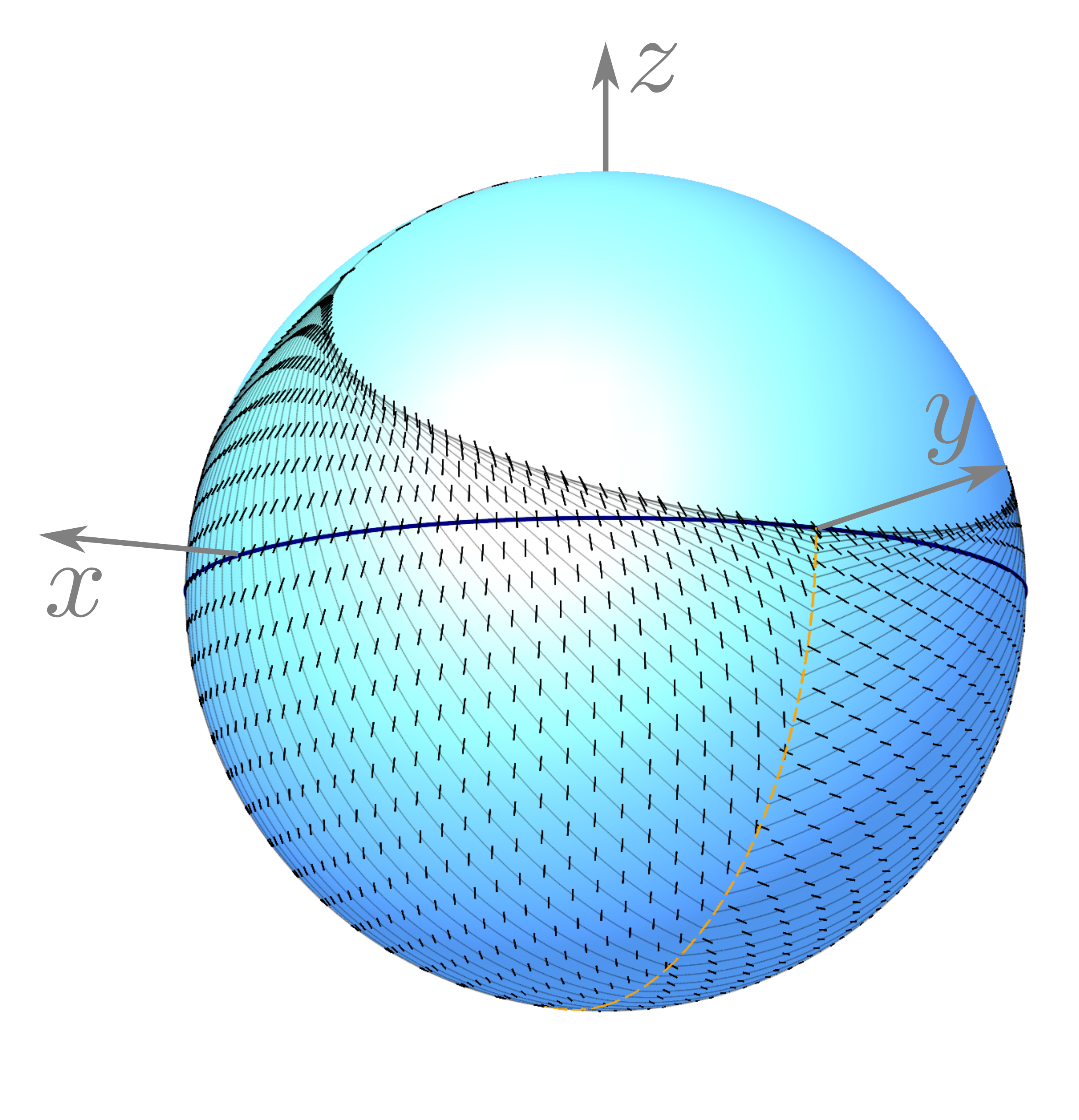}
\caption{$m=2$}
\end{subfigure}
\quad
\begin{subfigure}[b]{0.3\textwidth}
\centering
\includegraphics[width=\textwidth]{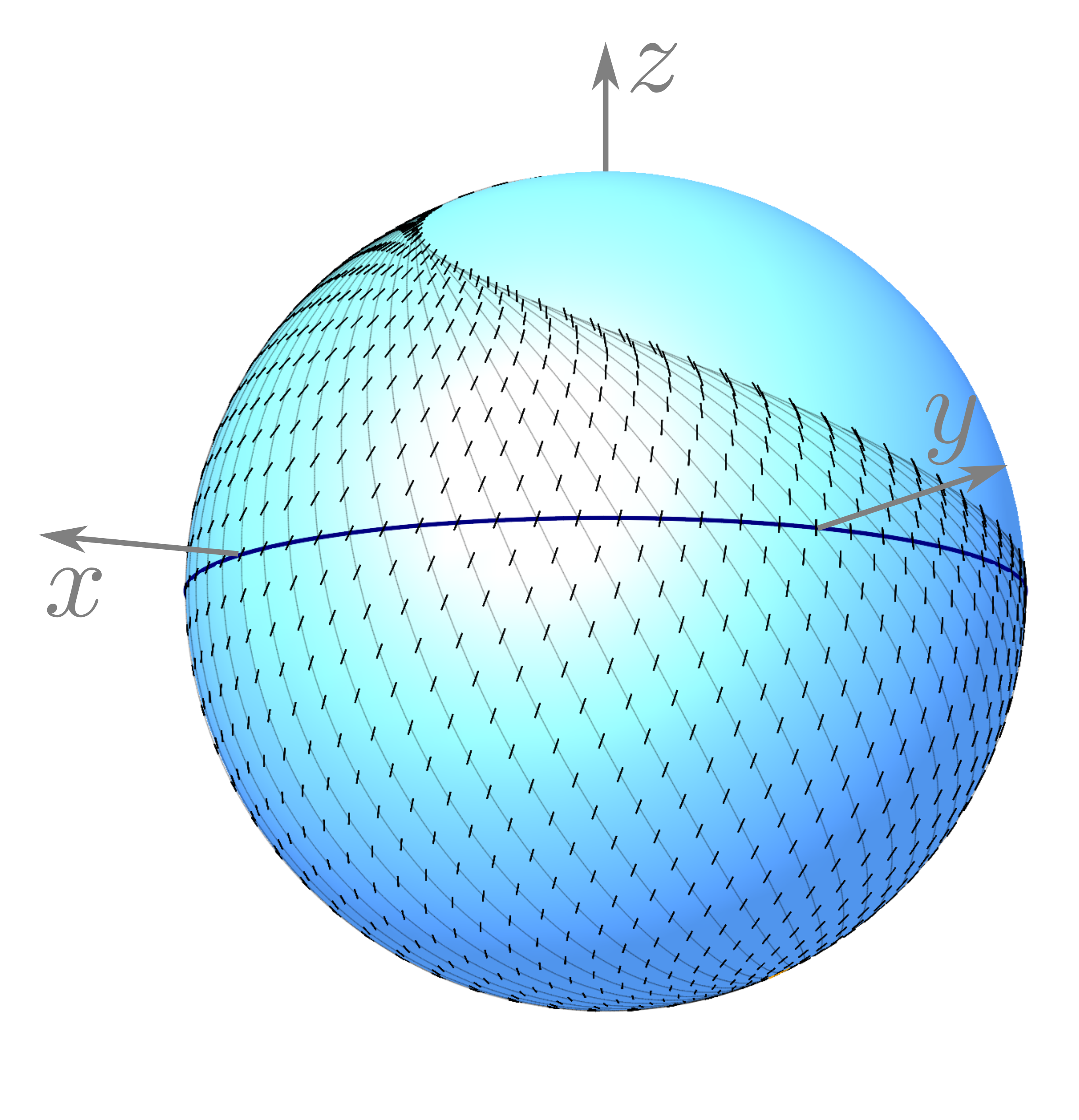}
\caption{$m=\frac32$}
\end{subfigure}
\caption{Local quasi-uniform distortions $\n$ that extend the Cauchy data \eref{eq:alpha_0_example} for $B=1$, $c_0=-\frac{\pi}{4}$, and three values of $m\neq1$, prescribed on the equator of the unit sphere (blue line). Directors are represented as headless vectors and the characteristic great circles that convey them are gray lines. The $2|m-1|$ points where $\tan\alpha_0=\frac1B$ belong to as many defect lines (dashed orange arcs).}
\label{fig:alpha_uniform}
\end{figure}

In Figure \ref{fig:alpha_uniform} are depicted examples of local extensions of the Cauchy data \eref{eq:alpha_0_example} for $c_0=-\arctan B$ and different values of $m$: they all cover parts of $\sphere$; each \emph{forbidden} point $p_i$  actually belongs to an entire defect line.

\subsection{Quasi-uniformity on a catenoid}\label{sec:catenoid}
The \emph{unit} catenoid is the  surface of revolution $\surface$ represented in a Cartesian frame $(\ex,\ey,\ez)$ via isothermal coordinates $(u,v)$ as
\begin{equation}\label{eq:catenoid_representation}
\rv(u,v) = \cosh v(\cos u\ex + \sin u\ey) + v\ez, 
\end{equation}
where $u\in[0,2\pi)$ and $v\in\R$.
Since
\begin{eqnarray}
\rvu &=& \cosh v(-\sin u\ex + \cos u\ey), \\
\rvv &=& \sinh v(\cos u\ex + \sin u\ey) + \ez, \\
r &=& \|\rvu\| = \|\rvv\| = \cosh v,
\end{eqnarray}
the corresponding moving frame $\framee$ is related to $\frameC$ through the equations
\begin{equation}\label{eq:iso_frame}
\system{
\eu &= -\sin u\ex + \cos u\ey, \\
\ev &= \tanh v(\cos u\ex + \sin u\ey) + \sech v\ez, \\
\vnu &= \sech v(\cos u\ex + \sin u\ey) - \tanh v\ez,
}
\end{equation}
and its connectors $(\cv,\vd_u,\vd_v)$ can easily be shown to be
\begin{equation}
\vc = -\sech v\tanh v\eu, \quad
\vdu = -\sech^2 v\eu, \quad
\vdv = \sech^2 v\ev.
\end{equation}
It follows from \eref{eq:K} that the Gaussian curvature of $\surface$ is
\begin{equation}
	\label{eq:K_catenoid}
	K=-\sech^4 v.
\end{equation}

We shall represent $\n$ as in \eref{eq:n_representation} and write the requirement of quasi-uniformity  in \eref{eq:quasi_pde} as 
\begin{equation}\label{eq:catenoid_pde}
\au(B\sin\alpha - \cos\alpha) - \av(B\cos\alpha + \sin\alpha) = \tanh v(B\sin\alpha - \cos\alpha).
\end{equation}
Letting 
\begin{equation}
	\label{eq:omega_definition}
\omega:=\alpha-\gamma = \alpha+\arctan B
\end{equation}
be the angle that the characteristic geodesic makes with $\eu$, we simplify \eref{eq:catenoid_pde} in the form
\begin{equation}\label{eq:catenoid_pde_beta}
\omega_{,u}\cos\omega + \omega_{,v}\sin\omega = \tanh v\cos\omega
\end{equation}
and we accordingly replace \eref{eq:lagrange_charpit} with the system
\begin{equation}\label{eq:catenoid_pde_system}
\system{
\dot{u} &= \cos\omega, \\
\dot{v} &= \sin\omega, \\
\dot\omega&= \tanh v\cos\omega.
}
\end{equation}  

\begin{figure}[h!]
	\centering
	\begin{subfigure}[b]{0.3\textwidth}
		\centering
		\includegraphics[width=\textwidth]{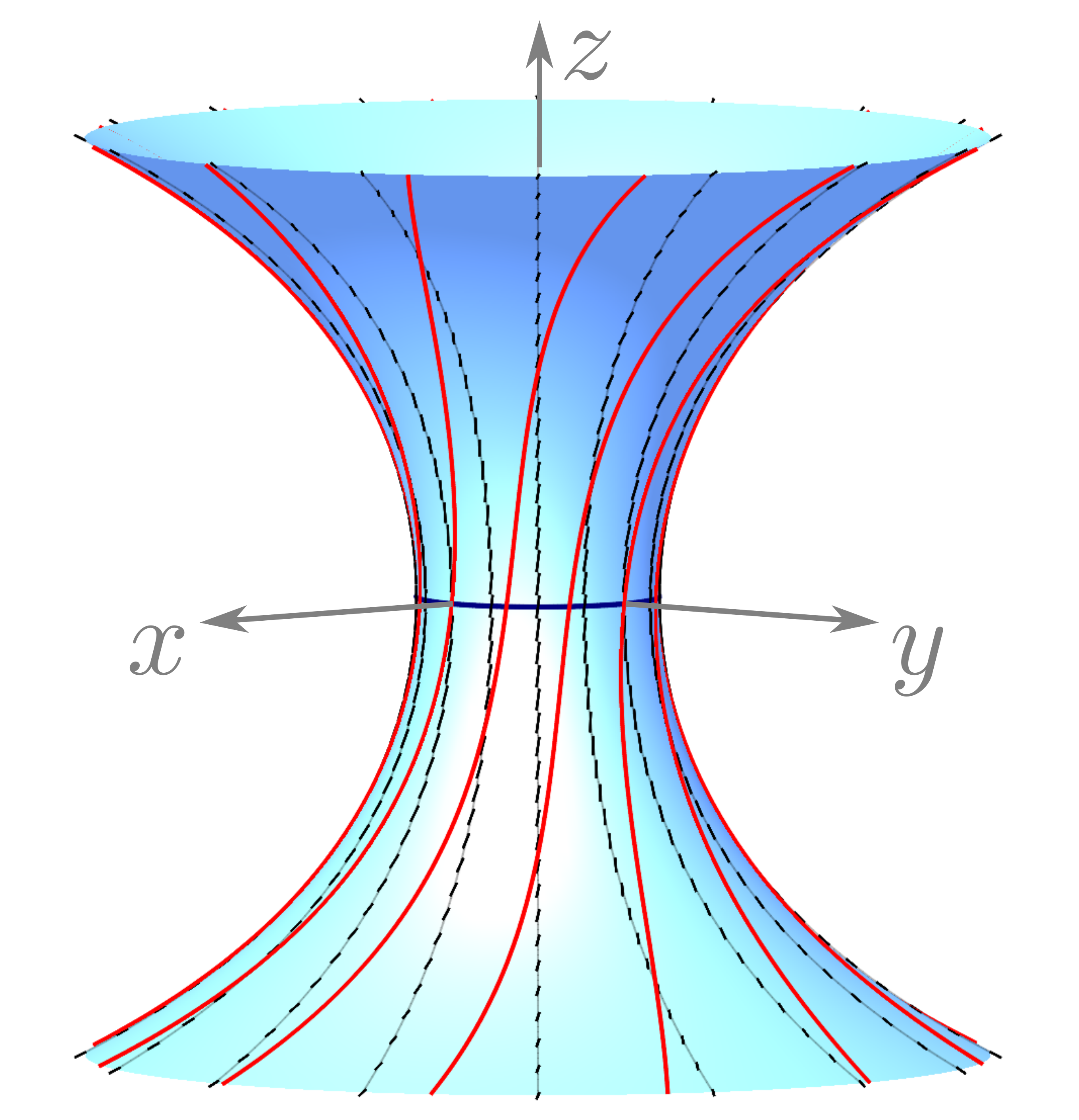}
		\caption{$B = \frac14$}
	\end{subfigure}
	$\quad$
	\begin{subfigure}[b]{0.3\textwidth}
		\centering
		\includegraphics[width=\textwidth]{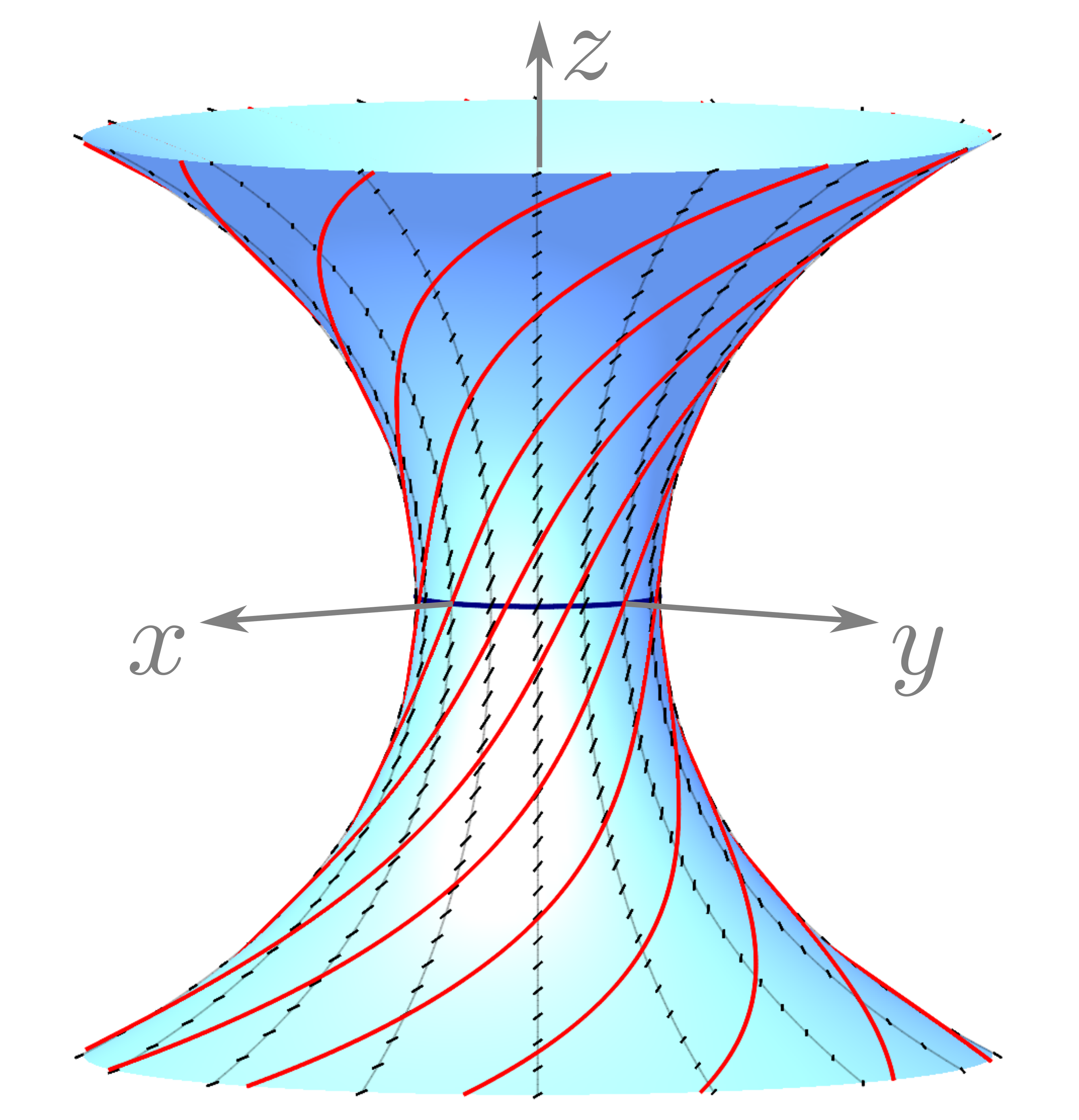}
		\caption{$B = 1$}
	\end{subfigure}
	$\quad$
	\begin{subfigure}[b]{0.3\textwidth}
		\centering
		\includegraphics[width=\textwidth]{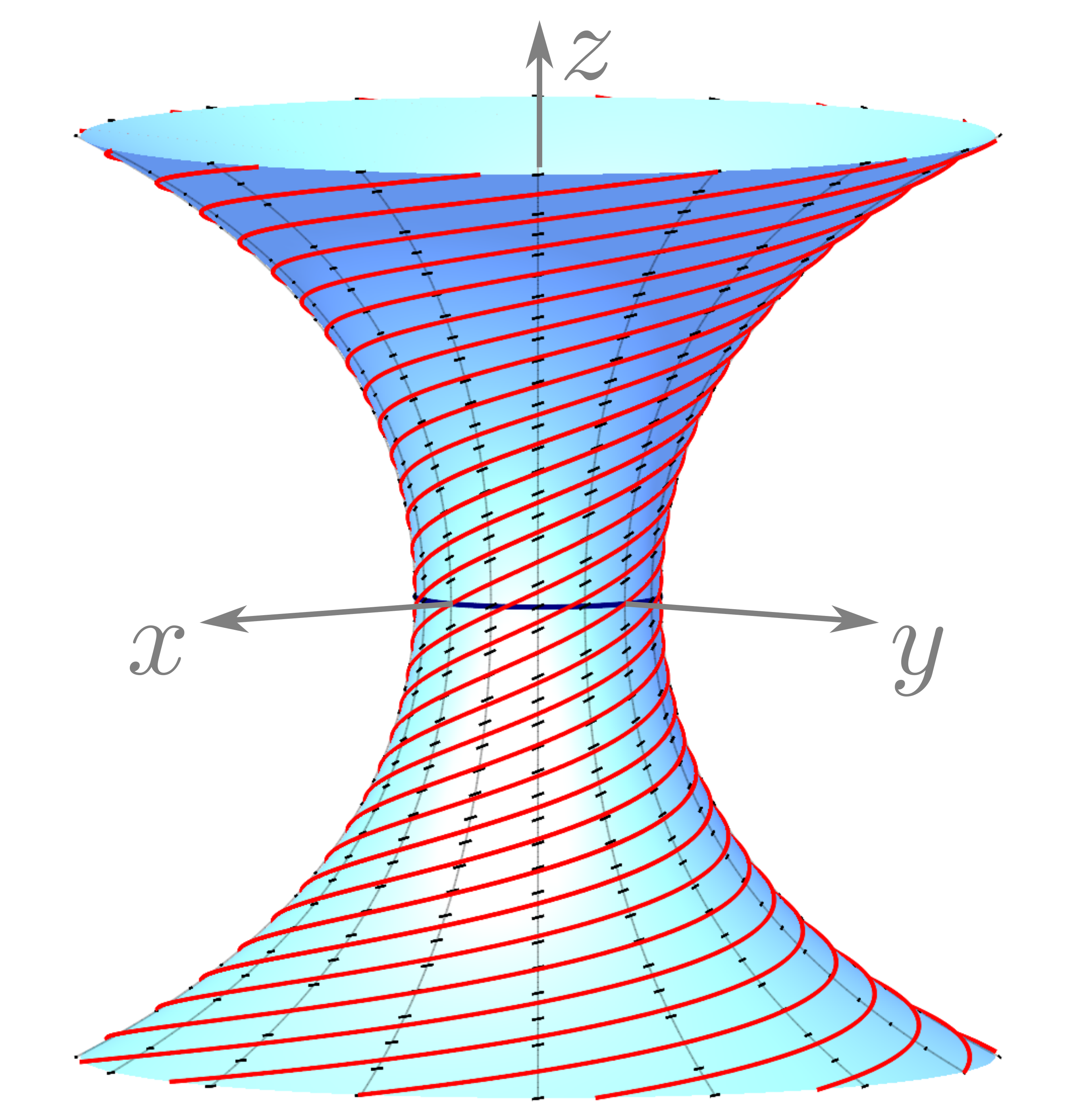}
		\caption{$B = 4$}
	\end{subfigure}
	\caption{Examples of global quasi-uniform distortions $\n$ whose characteristic geodesics are the meridians of the catenoid. Directors (represented as headless vectors) make the (constant) angle $\alpha = \frac\pi2 -\arctan B$ with the meridians.   The field lines of $\n$ (in red) are the loxodromes of the catenoid, paralleling those of the sphere shown in Figure~\ref{fig:alpha_constant}.}
	\label{fig:cat_meridians}
\end{figure}

A special solution of \eref{eq:catenoid_pde_system} is obtained for $\omega\equiv\frac{\pi}{2}$, corresponding to the system of meridians $u\equiv u_0$ (with arbitrary $u_0$). It readily follows from \eref{eq:omega_definition} that $\alpha\equiv\frac{\pi}{2}-\arctan B$ and the corresponding field lines of $\n$ are loxodromes  of the catenoid; these are \emph{global} quasi-uniform distortions paralleling the ones encountered for the sphere
(some examples are depicted in Figure~\ref{fig:cat_meridians}). 

More generally, since the meridians of a surface of revolution $\surface$ are geodesics, letting $\eu$ be oriented along the parallels of $\surface$, by use of \eref{eq:n_representation}, we can write the quasi-uniform distortion conveyed by the meridians of $\surface$ as
\begin{equation}
	\label{eq:n_meridians}
	\n=\frac{1}{\sqrt{B^2+1}}(B\eu+\ev),
\end{equation}
which agrees with \eref{eq:char_t_t_perp} (provided that we choose $\vt=\ev$ and $\vtp=-\eu$). The field lines of \eref{eq:n_meridians} are loxodromes to the meridians of $\surface$.

There are other global quasi-uniform distortions on the catenoid whose characteristic geodesics are \emph{not} meridians. For definiteness, we shall confine attention to solution of  \eref{eq:catenoid_pde_system} for which $0\le\omega\le\pi$. 

We start by singling out the geodesic $\curve$ passing through the point $(u_0,v_0)$ with slope $\omega_0$ over the local parallel ($v=v_0$). By integrating the equation
\begin{equation}
	\label{eq:catenoid_geodesic}
	\frac{\dd v}{\dd\omega}=\frac{\tan\omega}{\tanh v},
\end{equation}
which is an immediate consequence of  \eref{eq:catenoid_pde_system}, we obtain that
\begin{equation}\label{eq:catenoid_geodesic_solution}
	\cos\omega=\cos\omega_0\frac{\cosh v_0}{\cosh v},
\end{equation}
which, in particular, says that $\sgn(\cos\omega)=\sgn(\cos\omega_0)$. The geodesic $\curve$ then extends indefinitely away from the equator, as \eref{eq:catenoid_geodesic_solution} is compatible for all $v$ such that $|v|>|v_0|$. For $\curve$ to reach the equator, and thus be capable of conveying a global quasi-uniform distortion, it must be
\begin{equation}
	\label{eq:Omega_0}
	\Omega_0:=|\cos\omega_0|\cosh v_0<1.
\end{equation}
When this condition is met, by use of \eref{eq:catenoid_geodesic_solution}, we also derive from \eref{eq:catenoid_pde_system} the differential equation\footnote{The same equation can easily be reached by applying Clairaut's therem to \eref{eq:catenoid_representation} (see, for example, \S9.3 of \cite{pressley:elementary}).}
\begin{equation}
	\label{eq:catenoid_geodesic-clairaut}
	\frac{\dd u}{\dd v}=\frac{1}{\sqrt{\frac{\cosh^2v}{\Omega_0^2}-1}},
\end{equation}
from which it follows that
\begin{equation}\label{eq:catenoid_geodesic_representation}
u = u_0 + \Omega_0\bigg[\ellF\bigg(\frac1{\cosh v_0}, \frac{1}{\Omega_0}\bigg)
-\ellF\bigg(\frac1{\cosh v},  \frac{1}{\Omega_0}\bigg)\bigg],
\end{equation} 
where $\ellF$ is the incomplete elliptic integral of the first kind. As $v\to\pm\infty$, $\curve$ approaches two meridians separated by the angular distance
\begin{equation}
\Delta u =2\Omega_0\ellK(\Omega_0),
\end{equation}
where $\ellK$ is the complete elliptic integral of the first kind.

All other characteristic geodesics needed to construct the desired global quasi-uniform distortion can be obtained by rotating $\curve$ around the symmetry axis of the catenoid (which amounts to make $u_0$ span in \eref{eq:catenoid_geodesic_representation} the whole interval $[0,2\pi]$).

\begin{figure}[h!]
\centering
\begin{subfigure}[b]{0.3\textwidth}
\centering
\includegraphics[width=\textwidth]{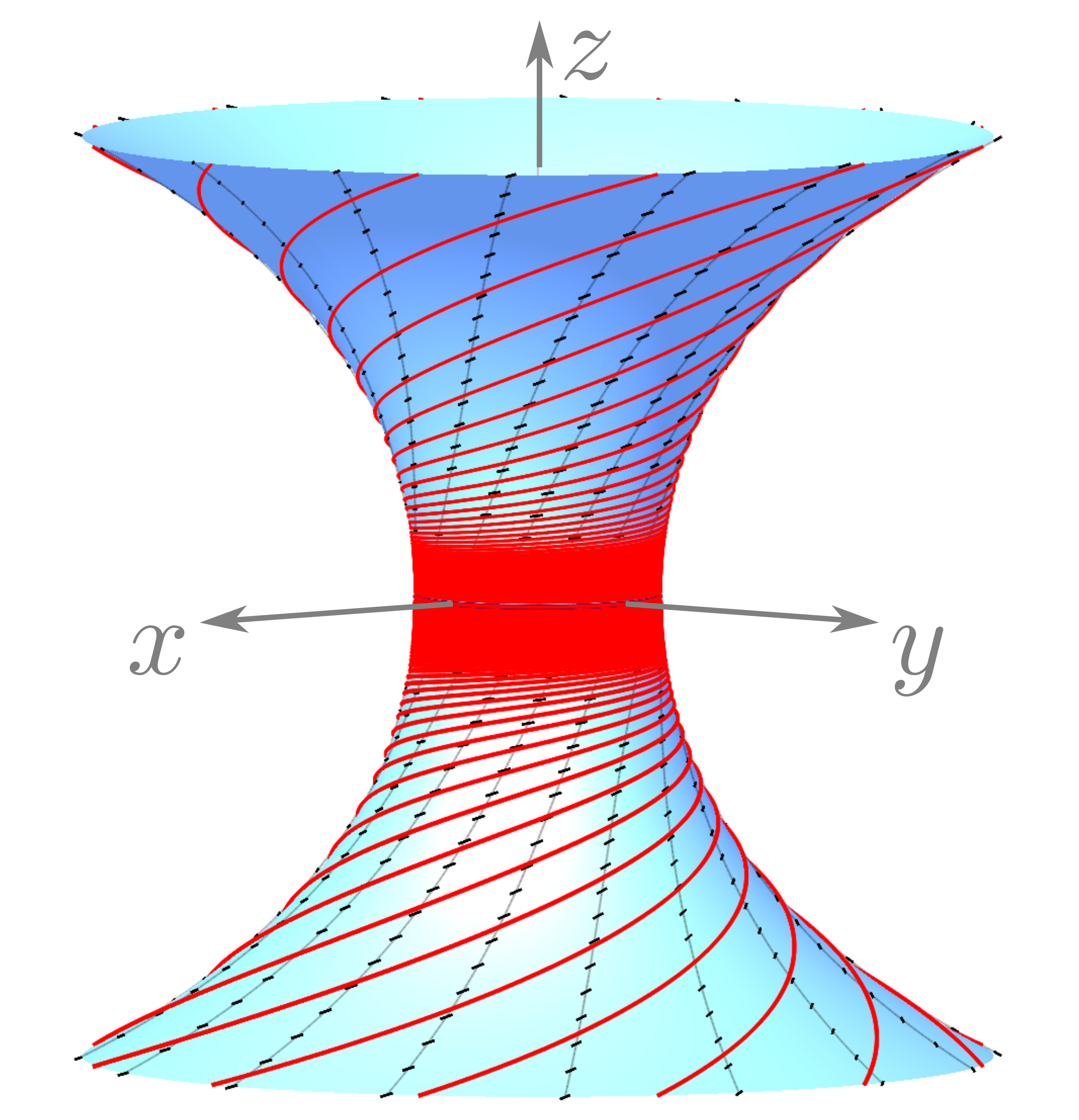}
\caption{$\alpha_0 = 0$}
\end{subfigure}
$\quad$
\begin{subfigure}[b]{0.3\textwidth}
\centering
\includegraphics[width=\textwidth]{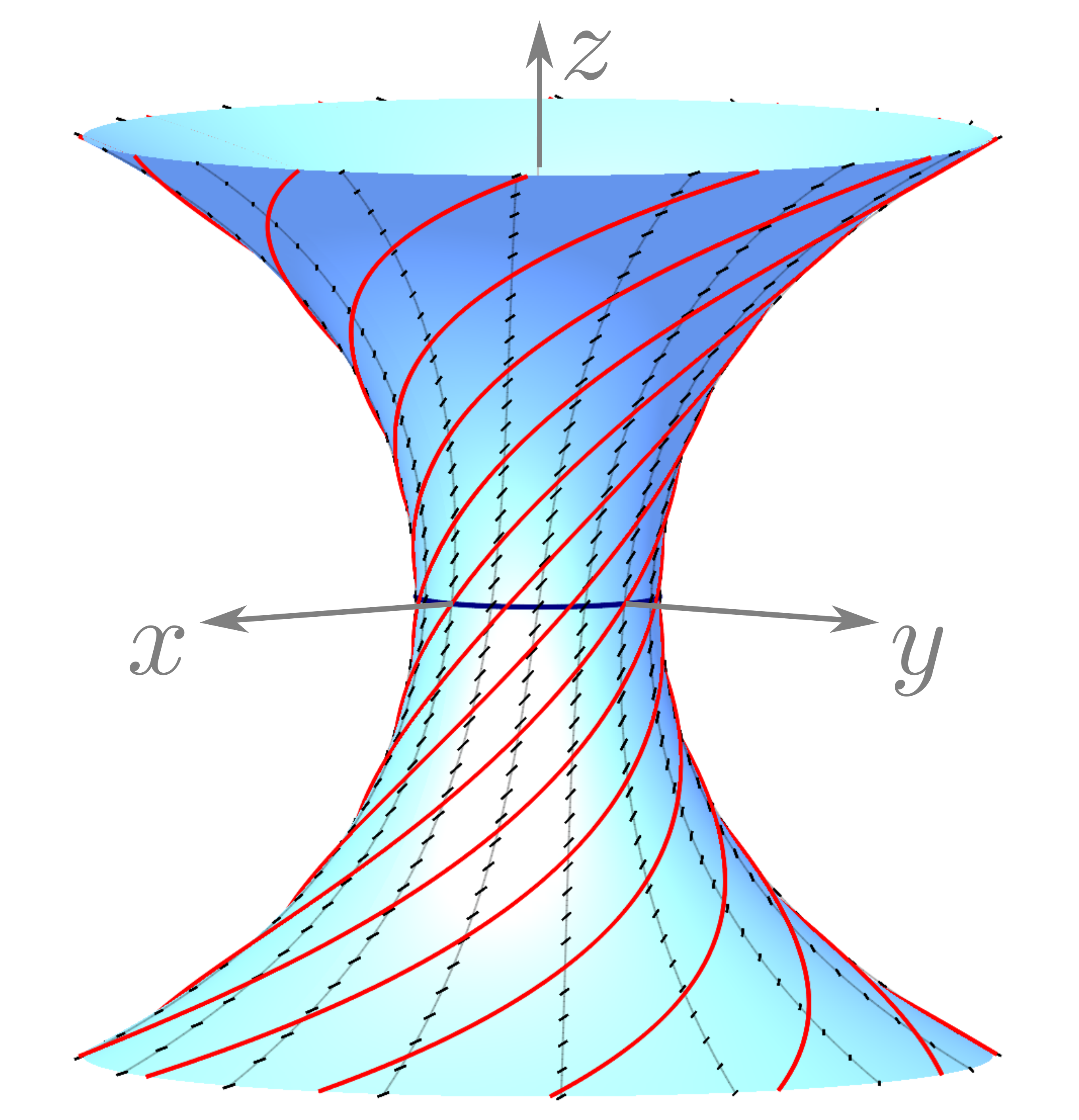}
\caption{$\alpha_0 = \frac\pi6$}
\end{subfigure}
$\quad$
\begin{subfigure}[b]{0.3\textwidth}
\centering
\includegraphics[width=\textwidth]{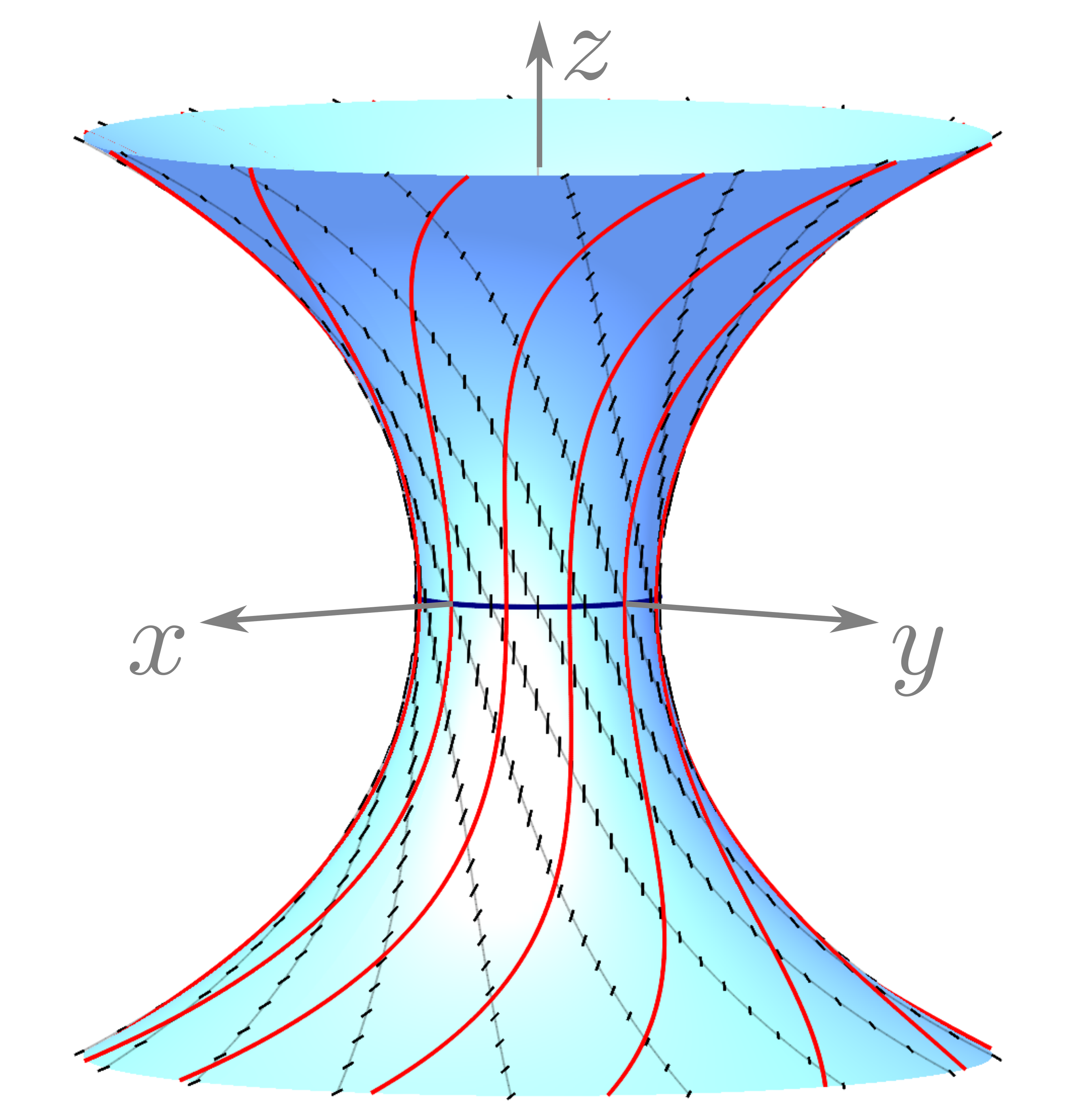}
\caption{$\alpha_0 = \frac\pi2$}
\end{subfigure}
\caption{Examples of global quasi-uniform distortions $\n$ conveyed by characteristic geodesics other than meridians. Here, $\alpha\equiv\alpha_0$ on the equator (blue line) and $\omega_0$ is determined according to  \eref{eq:catenoid_omega_0}, with $B=1$. The field lines of $\n$ are depicted in red, and the conveying geodesics in  gray. Directors are represented as headless vectors.}
\label{fig:cat_parallel}
\end{figure}

We proceed by choosing $v_0=0$ and setting $\alpha\equiv\alpha_0$ on the equator, so that, by \eref{eq:omega_definition} and \eref{eq:Omega_0}, we have that
\begin{equation}\label{eq:catenoid_omega_0}
\omega_0 = \alpha_0 + \arctan B\quad\textrm{and}\quad\Omega_0=\cos\omega_0.
\end{equation}
Examples of this construction are illustrated  in Figure \ref{fig:cat_parallel} for $B=1$ and three values of $\alpha_0$.

\begin{figure}[h!]
\centering
\begin{subfigure}[b]{0.3\textwidth}
\centering
\includegraphics[width=\textwidth]{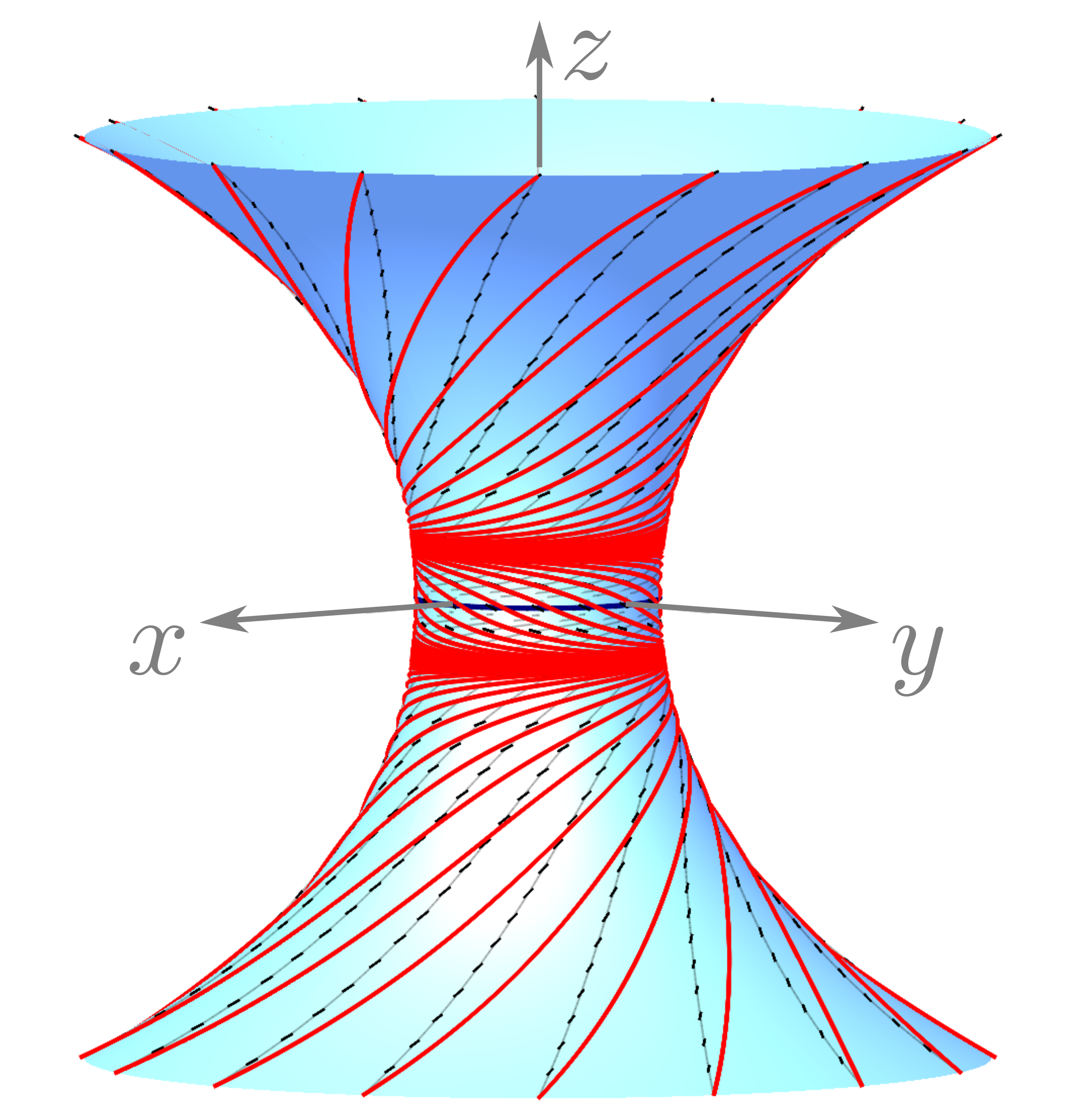}
\caption{$B = \frac14$}
\end{subfigure}
$\quad$
\begin{subfigure}[b]{0.3\textwidth}
\centering
\includegraphics[width=\textwidth]{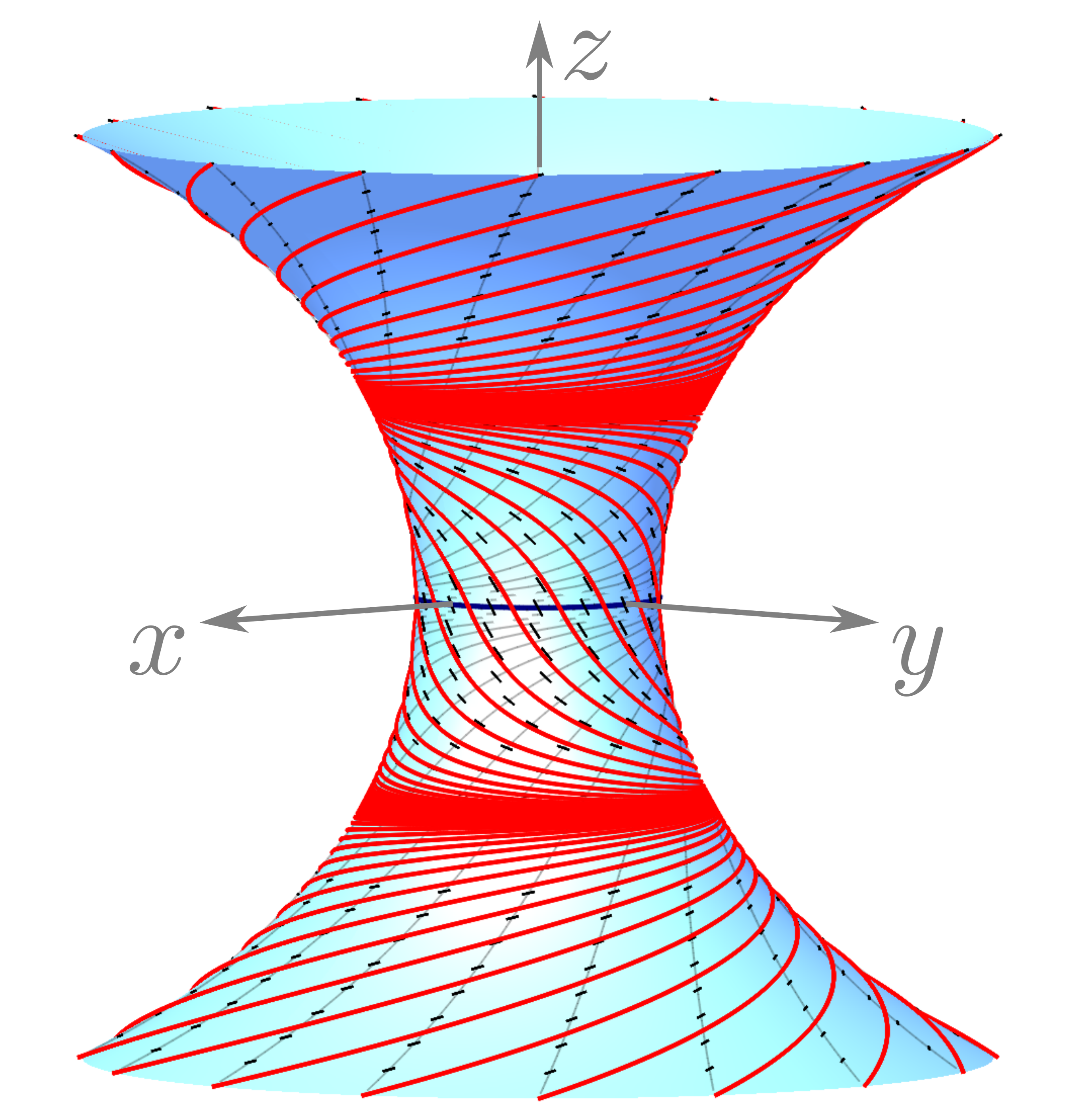}
\caption{$B = 1$}
\end{subfigure}
$\quad$
\begin{subfigure}[b]{0.3\textwidth}
\centering
\includegraphics[width=\textwidth]{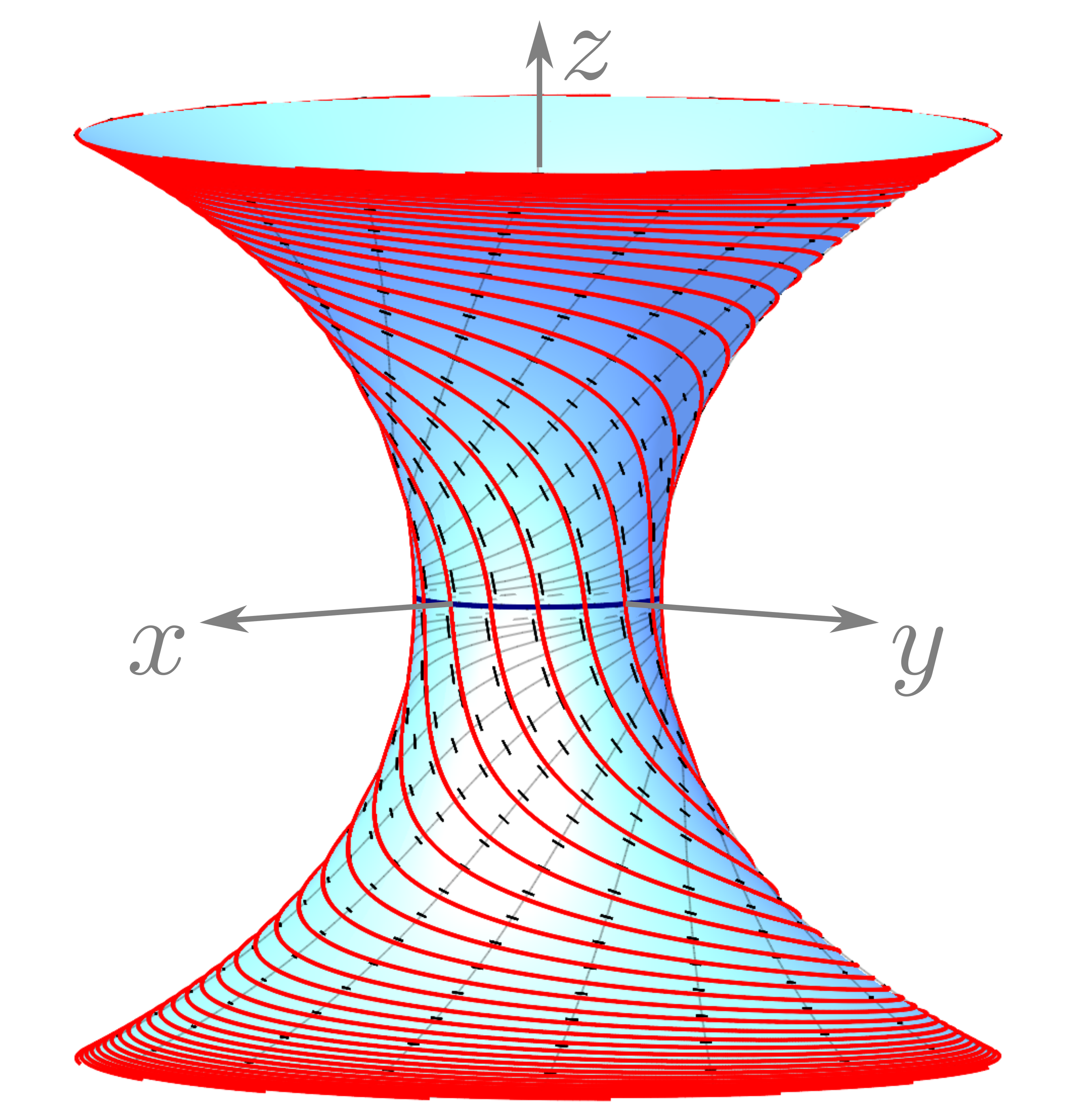}
\caption{$B = 4$}
\end{subfigure}
\caption{Examples of global quasi-uniform distortions $\n$ conveyed by characteristic geodesics that wind indefinitely closer and closer to the equator (itself a characteristic geodesic). Here, $\omega_0=0$ and, according to \eref{eq:omega_definition}, $\alpha_0=-\arctan B$. The conveying geodesics are gray and the  equator is blue. The two parallels  designated by \eref{eq:catenoid_parallels}, upon which the (red)  field lines of $\n$ are tangent and where they revert their sense of winding are placed symmetrically relative to the equator.}
\label{fig:cat_limit}
\end{figure}

A very special case arises if $\sin\omega_0=0$ (and $\Omega_0=1$): then the equator is itself a characteristic geodesic, as easily follows from \eref{eq:catenoid_pde_system}. In this case, \eref{eq:catenoid_geodesic-clairaut} remains valid for all characteristic geodesics other than the equator, but with $\Omega_0=1$; its integral is
\begin{equation}
	\label{eq:catenoid_special_case}
	u=u_0-2\arctanh(\mathrm{e}^{-|v|}),
\end{equation}
which represents characteristic geodesics that draw closer and closer to the equator, coiling around it infinitely many times while approaching it. Correspondingly, as shown in Figure~\ref{fig:cat_limit}, the  field lines of the conveyed quasi-uniform distortion $\n$ are smooth and become tangent to a parallel, where they revert their sense of winding, when $\sin\alpha=0$. By \eref{eq:omega_definition}, \eref{eq:catenoid_geodesic_solution}, and \eref{eq:Omega_0}, there  are two such parallels, identified by the roots $v$ of the following equation
\begin{equation}\label{eq:catenoid_parallels}
\sin\Bigg(\arccos\frac1{\cosh v}-\arctan B\Bigg)=0.
\end{equation}

\section{Conclusions}\label{sec:colclusion}
The need to categorize nematic fields on surfaces has prompted this study. The simplest category one can think of, which may be relevant to the description on pure geometric grounds of possible reference states of a general elastic theory, is that of \emph{uniform} distortions, whose invariant characteristics are \emph{constant}. However, only special surfaces can host such fields, those with constant \emph{negative} Gaussian curvature, or \emph{pseudospherical} surfaces, in Beltrami's language.

For a generic surface, we need a milder notion to make room for other significant fields. We found that notion to be \emph{quasi-uniformity}, which requires the distortion characteristics to be in constant \emph{ratio} to one another, instead of merely constant.

We proved that quasi-uniform nematic distortions are generated by parallel transporting (in Levi-Civita's sense) the director $\n$ along geodesics. This general construction applies to all (regular) surfaces, irrespective of their Gaussian curvature, and gives rise to a plethora of director fields, which we have only started to explore. These include (but are not limited to) global director fields whose field lines are loxodromes to the meridians of surfaces of revolution. A catenoid, for one, has also global quasi-uniform distortions conveyed by geodesics other than the meridians.

By the theorem proved in this paper, (geodesically) complete surfaces\footnote{These are surfaces whose geodesics can be mapped to the whole of $\R$.} present themselves as the most natural setting where to seek global quasi-uniform distortions. This quest just began here.

Quasi-uniform distortions proved also to be a means to extend locally a nematic field prescribed along a curve $\curve$ on  a surface $\surface$. Here the issue is how a uniform field on $\curve$ can be extended quasi-uniformly on $\surface$. The geometric construction that we provided has been applied to a number of illustrative cases on a sphere. The extension is always unique, but seldom global.

We also introduced the \emph{character} of a surface quasi-uniform field. This is a scalar function that expresses the invariant distortions of the field in terms of some reference constant measures; it is defined within a sign and solves an evolution equation along the conveying geodesics. We have determined it for some of the illustrative cases considered here, but a systematic study is lacking (and would be desirable).

This paper has perhaps gathered enough evidence to prompt a more systematic inquiry on quasi-uniform nematic distortions on surfaces, as the complexity of their geometric structure may be indicative of a significance yet to be fully appreciated.

\ack
{Both authors are members of \emph{GNFM}, a branch of \emph{INdAM}, the Italian Institute for Advanced Mathematics. A.P. wishes to acknowledge financial support from the Italian MIUR through European Programmes REACT EU 2014-2020 and PON 2014-2020 CCI2014IT16M2OP005.}

\appendix

\section{Surface gradient decompositions}\label{app:surface_gradient}
Let $\surf$ be a smooth surface (at least of class $C^2$) oriented by the unit normal $\vnu$ and let $\n:\surf\to\sphere$ be a (locally) differentiable unit vector field such that $\n\cdot\vnu=0$ everywhere on $\surf$.

By differentiating over $\surface$ the identities $\n\cdot\n\equiv1$ and $\n\cdot\vnu\equiv0$, we easily see that
\begin{equation}\label{eq:gradnt_nu} 
(\grads\n)^\top\n=\bm{0}\quad\textrm{and}\quad
(\grads\n)^\top\vnu = -(\grads\vnu)^\top\n, 
\end{equation}
the latter of which is equivalent to \eref{eq:swap} in the main text, as $\curvature$ is a symmetric tensor.
Thus, it follows from \eref{eq:curls} and \eref{eq:b_S_T_surface} that the surface bend vector can also be written as
\begin{equation}\label{eq:bends_bis}
\bends = -\curls\n\times\n = -(\grads\n)\n.
\end{equation}

Since $(\grads\n)\normal\equiv\bm{0}$, to decompose $\grads\n$ completely we further need only know how it acts on $\np$. For this reason, we project $(\grads\n)\np$ along the unit vectors of the moving frame $\framen$, as, by the first equation in \eref{eq:gradnt_nu}, its projection along $\n$ vanishes. Since the (three-dimensional) identity $\id$ can also be written as
 \begin{equation}
 	\label{eq:identity}
 	\id=\n\otimes\n+\np\otimes\np+\normal\otimes\normal,
 \end{equation}
we have that\footnote{Here the inner product $\mathbf{A}\cdot\mathbf{B}$ of tensors $\mathbf{A}$ and $\mathbf{B}$ is defined as $\mathbf{A}\cdot\mathbf{B}:=\tr(\mathbf{A}^\top\mathbf{B})$.}
\begin{equation}\label{eq:np_gradn_np}
\np\cdot(\grads\n)\np
= \n\cdot(\grads\n)\n + \np\cdot(\grads\n)\np + \vnu\cdot(\grads\n)\vnu= \id\cdot\grads\n=\dvs\n.
\end{equation}
Moreover, again by \eref{eq:curls}, 
\begin{equation}
\label{eq:nu_gradn_n}
\vnu\cdot(\grads\n)\np
= \vnu\cdot(\grads\n)\np - (\grads\n)\vnu\cdot\np 
= \vnu\cdot\big[\grads\n - (\grads\n)^\top\big]\np
=\normal\cdot\curls\n\times\np = \n\cdot\curls\n,
\end{equation}
where use has also been made of the identity 
\begin{equation}
	\label{eq:mixed_product_symmetries}
	\bm{a}\cdot\bm{b}\times\bm{c}=\bm{b}\cdot\bm{c}\times\bm{a}=\bm{c}\cdot\bm{a}\times\bm{b},
\end{equation}
valid for any vectors $\bm{a}$, $\bm{b}$, $\bm{c}$. Combining \eref{eq:np_gradn_np} and \eref{eq:nu_gradn_n}, we arrive at
\begin{equation}\label{eq:gradn_np}
(\grads\n)\np =  (\dvs\n)\np + (\n\cdot\curls\n)\vnu.
\end{equation}
Putting together \eref{eq:bends_bis} and \eref{eq:gradn_np}, we thus prove \eref{eq:surface_gradient_representation} in the main text, which can easily be rewritten as in \eref{eq:surf_dec}, if one so wishes.

We now prove the decomposition of $\curls\n$ in \eref{eq:surface_curl_decomposition}. By \eref{eq:mixed_product_symmetries} and \eref{eq:curls}, we have that 
\begin{equation}
	\label{eq:curlsn_n}
\curls\n\cdot\n
=-\curls\n\cdot\normal\times\np=-\np\cdot\curls\n\times\normal
=\np\cdot(\grads\n)^\top\normal
=-\np\cdot\curvature\n,
\end{equation}
where also \eref{eq:gradnt_nu} has been used. Similarly, we get 
\begin{equation}
	\label{eq:curlsn_np}
\curls\n\cdot\np
=
\curls\n\cdot\normal\times\n=\n\cdot\curls\n\times\normal=-\n\cdot(\grads\n)^\top\normal
\n\cdot\curvature\n
\end{equation}
and
\begin{equation}
	\label{eq:curlsn_nu}
\curls\n\cdot\normal
=
\curls\n\cdot\n\times\np=\np\cdot\curls\n\times\n
=\np\cdot(\grads\n)\n.
\end{equation}
Equations \eref{eq:curlsn_n}, \eref{eq:curlsn_np}, and \eref{eq:curlsn_nu} collectively prove \eref{eq:surface_curl_decomposition}.

\section{Non-parallel transport}\label{app:transports}
We find it instructive to show in this appendix that the field \eref{eq:sphere_n_global}, which is parallel transported along meridians is neither parallel nor uniformly transported along another generic geodesic of $\sphere$. To this end, we represent $\n$ as in \eref{eq:field}
with $\alpha$ constant, and consider a generic great circle $\great$ of $\sphere$, different from both equator and meridians. With no loss of generality, this can be obtained by rotating the equator around the $x$ axis by a (constant) angle $\beta\in\big(0,\frac\pi2\big)$. By \eref{eq:spherical}, $\great$ is then represented by the curve
\begin{equation}\label{eq:great_circle}
\eqalign{
\vp(\sigma) &:= \big[\id + \sin\beta{\tW}(\ex) + (1-\cos\beta){\tW}^2(\ex)\big](\cos\sigma\ex + \sin\sigma\ey) \\
&\,= \cos\sigma\ex + \cos\beta\sin\sigma\ey + \sin\beta\sin\sigma\ez,
}
\end{equation}
where $\sigma\in[0,2\pi)$ is the arc-length parameter.\footnote{Equation \eref{eq:great_circle} easily follows from Euler's representation of the rotation $\mathbf{R}(\beta)$ of angle $\beta$ about an axis $\bm{e}$,
$$
\mathbf{R}(\beta)=\id+\sin\beta\W(\bm{e})+(1-\cos\beta)\W^2(\bm{e}),
$$
where $\W(\bm{e})$ is the skew-symmetric tensor with axial vector $\bm{e}$.} The unit vector $\vp(\sigma)$ can be made coincide with $\er$ in the moving frame $(\ep,\et,\er)$ by letting the azimuthal angle $\vartheta$ and the polar angle $\varphi$ depend on $\sigma$ as
\begin{equation}\label{eq:sigma_parametrization}
\vartheta(\sigma) = \arctan(\cos\beta\tan\sigma)
\quad\textrm{and}\quad
\varphi(\sigma) = \arccos(\sin\beta\sin\sigma).
\end{equation}

By use of \eref{eq:sigma_parametrization} in \eref{eq:spherical}, we obtain from \eref{eq:field} that 
\begin{equation}
\n' 
= \cos\alpha\ep' + \sin\alpha\et'
= \cos\alpha(-\varphi'\er + \vartheta'\cos\varphi\et)
- \vartheta'\sin\alpha(\sin\varphi\er+ \cos\varphi\ep),
\end{equation}
where a prime stands for differentiation with respect to $\sigma$. Since
\begin{equation}
\np = \er\times\n = - \sin\alpha\ep + \cos\alpha\et,
\end{equation}
and the geodesic curvature $\kappa_g$ of $\great$ vanishes, by \eref{eq:n_prime}, we have that
\begin{equation}\label{eq:alpha_great}
\gamma' = \n'\cdot\np = \vartheta'\cos\varphi 
= \frac{\cos\beta\sin\beta\sin\sigma}{\cos^2\sigma+ \cos^2\beta\sin^2\sigma},
\end{equation}
where \eref{eq:sigma_parametrization} has also been used. Integrating \eref{eq:alpha_great} we arrive at
\begin{equation}
	\label{eq:gamma_solution}
	\gamma=\gamma_0-\arctan(\tan\beta\cos\sigma),
\end{equation}
where $\gamma_0$ is an arbitrary constant. Since, for $\beta\in\big(0,\frac\pi2\big)$, $\gamma$ fails to be constant, we conclude that $\n$ in \eref{eq:field} is \emph{not} parallel transported along $\great$. For $\beta=0$,  \eref{eq:gamma_solution} implies that $\gamma=\gamma_0$, and so $\n$ is  parallel transported along the equator $\curve_0$.

\section{Surfaces of revolution}\label{sec:revolution}
In this appendix, we determine the general quasi-uniformity \emph{character} of $\n$ when this is conveyed by a special system of geodesics that all surfaces of revolution possess, independently of their shape, that is, the meridians.

Let the surface $\surface$ be represented in a Cartesian frame $\frameC$ by the mapping
\begin{equation}
	\label{eq:revolution_r}
	\rv(u,v)=\psi(v)(\cos u\ex+\sin u\ey)+\phi(v)\ez,
\end{equation}
where $(u,v)\in[0,2\pi)\times I$ (with $I\subseteq\mathbb{R}$) and $\psi$, $\phi$ are smooth functions of $v$, the former of which is assumed to be positive.\footnote{The parameterization in \eref{eq:revolution_r} is not necessarily isothermal.} $\surface$ is a surface of revolution about $\ez$: its cut with the plane $z=\phi(v)$ is a circle of radius $\psi(v)$. The meridians of $\surface$ are the lines with $u$ constant.

Elementary calculations show that the moving frame associated with coordinates $(u,v)$ is given by
\begin{equation}
	\label{eq:revolution_frame}
	\system{
		\eu&=-\sin u\ex+\cos u\ey,\\
		\ev&=\frac{1}{\sqrt{\psi'^2+\phi'^2}}\{\psi'(\cos u\ex+\sin u\ey)+\phi'\ez\},\\
		\normal&=\frac{1}{\sqrt{\psi'^2+\phi'^2}}\{\phi'(\cos u\ex+\sin u\ey)-\psi'\ez\},
	}
\end{equation}
where $\normal=\eu\times\ev$ and a prime $'$ denotes differentiation with respect to $v$. For a generic curve $\curve$ on $\surface$ parameterized as $t\mapsto\rv(u(t),v(t))$, we easily find that
\begin{equation}
	\label{eq:revolution_r_dot}
	\dot{\rv}=\dot{u}\psi\eu+\dot{v}\sqrt{\psi'^2+\phi'^2}\ev,
\end{equation}
where a superimposed dot $\dot{\null}$ denotes differentiation with respect to the parameter $t$. It follows from \eref{eq:revolution_r_dot} that for a differentiable function $\chi=\chi(u,v)$,
\begin{equation}
	\label{eq:revolution_grad_chi}
	\grads\chi=\frac{1}{\psi}\partial_u\chi\eu+\frac{1}{\sqrt{\psi'^2+\phi'^2}}\partial_v\chi\ev.
\end{equation}

Similarly, by differentiating along $\curve$ the moving frame $\framee$, we conclude that the gliding laws \eref{eq:gliding_laws} apply with connectors given by
\begin{equation}
	\label{eq:revolution_connectors}
	\system{
		\cv&=-\frac{\psi'}{\psi\sqrt{\psi'^2+\phi'^2}}\eu,\\
		\dvu&=-\frac{\phi'}{\psi\sqrt{\psi'^2+\phi'^2}}\eu,\\
		\dvv&=\frac{1}{\psi'^2+\phi'^2}\left[\phi'\left(\frac{\psi'}{\sqrt{\psi'^2+\phi'^2}}\right)'-\psi'\left(\frac{\phi'}{\sqrt{\phi'^2+\phi'^2}}\right)'\right]\ev.
	}
\end{equation}
In particular, it follows from \eref{eq:revolution_connectors} and \eref{eq:K} that 
\begin{equation}
	\label{eq:revolution_K}
	K=-\frac{\phi'}{\psi(\psi'^2+\phi'^2)^{3/2}}\left[\phi'\left(\frac{\psi'}{\sqrt{\psi'^2+\phi'^2}}\right)'-\psi'\left(\frac{\phi'}{\sqrt{\psi'^2+\phi'^2}}\right)'\right].
\end{equation}
By \eref{eq:character_evolution} and \eref{eq:revolution_grad_chi}, the quasi-uniform fields conveyed by the meridians of $\surface$ have a character $f$ that obeys the differential equation
\begin{equation}
	\label{eq:revolution_character_equation}
	f^2\pm\frac{f'}{\sqrt{\psi'^2+\phi'^2}}+K=0,
\end{equation}
where $K$ is given by \eref{eq:revolution_K}.\footnote{The reader is reminded that in \eref{eq:revolution_character_equation} a prime denote differentiation with respect to $v$, whereas in \eref{eq:character_evolution} it denotes differentiation with respect to arc-length.}

We now apply \eref{eq:revolution_character_equation} to the two specific cases considered in the main body of the paper, that is, for $\surface$ either a sphere or a catenoid. In the former case, identifying $u$ with $\vartheta$ and $v$ with $\frac{\pi}{2}-\varphi$,\footnote{While $(\vartheta,\varphi)$ are the same angles used in the main text (and illustrated in Figure~\ref{fig:polar}), the choice for $(u,v)$ made here differs from that in \eref{eq:sphere_iso}. But this will not affect our conclusions.} we can set
\begin{equation}
	\label{eq:revolution_sphere}
	\phi(v)=\sin v,\quad \psi(v)=\cos v,
\end{equation}
so that by \eref{eq:revolution_K} $K\equiv1$ and \eref{eq:revolution_character_equation} becomes
\begin{equation}
	\label{eq:revolution_character_equation_sphere}
	f^2\pm f'+1=0.
\end{equation}
Solving this equation subject to $f(0)=0$, we arrive at 
\begin{equation}
	\label{eq:revolution_character_sphere}
	f(v)=\pm\tan v=\pm\cot\varphi,
\end{equation}
which agrees with \eref{eq:representation_formulae}, provided that we set 
\begin{equation}
	\label{eq:revolution_S_0_B_0}
	S_0=\pm\frac{1}{\sqrt{B^2+1}}\quad\textrm{and}\quad B_0=\pm\frac{B}{\sqrt{B^2+1}}.
\end{equation}

Similarly, for the catenoid represented by \eref{eq:catenoid_representation}, we can set
\begin{equation}
	\label{eq:revolution_catenoid}
	\phi(v)=v,\quad\psi(v)=\cosh v,
\end{equation}
so that, by \eref{eq:K_catenoid}, \eref{eq:revolution_character_equation} becomes
\begin{equation}
	\label{eq:revolution_character_equation_catenoid}
	f^2\pm\sech vf'-\sech^4v=0.
\end{equation}
The solution of this equation subject to $f(0)=0$ is
\begin{equation}
	\label{eq:revolution_character_catenoid}
	f(v)=\pm\sech v\tanh v,
\end{equation}
which provides the quasi-uniformity character of the director field \eref{eq:n_meridians} conveyed by the meridians of a catenoid.


%

\end{document}